\documentclass[twocolumn]{aastex63}
\usepackage{xcolor}
\usepackage{hyperref}
\usepackage{enumerate}
\usepackage [caption=false]{subfig}
\usepackage{natbib}
\usepackage{url}
\usepackage{amsthm,amsmath,amssymb}

\usepackage{mathrsfs}
\newcommand\Nu{NuSTAR }
\newcommand\ec{$E_{\rm cut} $ }

\begin{document}

\title{The X-ray coronae in NuSTAR bright active galactic nuclei}
\correspondingauthor{Jia-Lai Kang \& Jun-Xian Wang} \email{ericofk@mail.ustc.edu.cn, jxw@ustc.edu.cn}

\author{Jia-Lai Kang} 
\author{Jun-Xian Wang}
\affiliation{CAS Key Laboratory for Research in Galaxies and Cosmology, Department of Astronomy, University of Science and Technology of
China, Hefei, Anhui 230026, China; ericofk@mail.ustc.edu.cn, jxw@ustc.edu.cn}
\affiliation{School of Astronomy and Space Science, University of Science and Technology of China, Hefei 230026, China}

\begin{abstract}
We present systematic and uniform analysis of \Nu data with 10--78 keV S/N $>$ 50, of a sample of 60 SWIFT BAT selected AGNs, 10 of which are radio-loud.
We measure their high energy cutoff \ec or coronal temperature $T_{\rm e}$ using three different spectral models to fit their NuSTAR spectra, and show a threshold in NuSTAR spectral S/N is essential for such measurements.
High energy spectral breaks are detected in the majority of the sample, and for the rest 
strong constraints to \ec or $T_{\rm e}$ are obtained. Strikingly, we find extraordinarily large \ec lower limits ($> 400$ keV, up to $>$ 800 keV) in 10 radio-quiet sources, whereas none in the radio-loud sample. Consequently and surprisingly, we find significantly larger mean $E_{\rm cut}$/$T_{\rm e}$ of radio-quiet sources compared with radio-loud ones.
The reliability of these measurements are carefully inspected and verified with simulations. 
We find a strong positive correlation between \ec and photon index $\Gamma$, which can not be attributed to the parameter degeneracy. 
The strong dependence of \ec on $\Gamma$, which could fully account for the discrepancy of \ec distribution between radio-loud and radio-quiet sources, indicates the X-ray coronae in AGNs with steeper hard X-ray spectra have on average higher temperature and thus smaller opacity.
However, no prominent correlation is found between \ec and $\lambda_{\rm edd}$.
In the $l$--$\Theta$ diagram, we find a considerable fraction of sources lie beyond the boundaries of forbidden regions due to runaway pair production, posing (stronger) challenges to various (flat) coronal geometries. 
\end{abstract}


\section{Introduction} \label{sec:intro}
\par The generally accepted disc-corona paradigm illustrates that the powerful hard X-ray emission universally found in active galactic nuclei (AGNs) is produced in the so-called corona \citep[e.g.,][]{Haardt_1991, Haardt_1993}. In this scenario the UV/optical photons from the accretion disk are upscattered to X-ray band through inverse Compton process by the hot electrons in the corona. However the physical nature of the corona remains yet unclear.
Particular matters of concern, for instance, include the location and geometry of the corona \citep{Fabian_2009, Alston_2020}, the underlying mechanism for X-ray spectra variability in individual sources \citep[e.g.][]{Wu_2020}, potential interactions within the corona like pair-production \citep{Fabian_2015}, and the relation between coronal and blackhole properties \citep{Ricci_2018, Hinkle_2021}. 

\par One of the most fundamental physical parameters of the corona is the temperature $kT_{\rm e}$.
The typical X-ray spectrum produced by the inverse Compton scattering within the corona is a power-law continuum, with a high energy cutoff. Such a cutoff ($E_{\rm cut}$) is a direct indicator of the coronal temperature, with \ec $\sim$ 2 $kT_{\rm e}$ or 3 $kT_{\rm e}$ for an optically thin or thick corona \citep{Petrucci_2001}. The Nuclear Spectroscopic Telescope Array \citep[NuSTAR;][]{Harrison_2013} is the
first hard X-ray telescope with direct-imaging capability above 10 keV. With its broad spectral coverage of 3--78 keV, \Nu\ has enabled the measurements (or lower limits) of $E_{\rm cut}$/$kT_{\rm e}$ in a number of AGNs 
\citep[e.g., ][]{Ballantyne_2014, Matt_2015, Ursini_2016, Kamraj_2018, Tortosa_2018, Molina_2019, Rani_2019, Panagiotou_2020, Porquet_2021, Hinkle_2021, Akylas_2021, Kamraj_2022}. Meanwhile, variations of $E_{\rm cut}$/$T_{\rm e}$ are also reported in a few individual sources \citep[e.g.,][]{Keek_2016, Zhangjx2018, Kang_2021}. 

\par However, even with \Nu spectra, the measurements of $E_{\rm cut}$/$kT_{\rm e}$ are highly challenging for most AGNs, primarily due to the limited spectral quality at the high energy end.
In many sample studies, only poorly constrained lower limits could be obtained for the dominant fraction of sources in the samples \citep[e.g.][]{Ricci_2018, Kamraj_2018, Panagiotou_2020, Kamraj_2022}, hindering further reliable statistical studies, e.g., to probe the dependence of $E_{\rm cut}$/$kT_{\rm e}$ on other physical parameters.
Meanwhile, the \ec measurements are often sensitive to the choice of spectral models. From this perspective, it is essential to perform uniform spectral fitting to a statistical sample with various models adopted.

\par Recently, we uniformly analyzed the \Nu spectra for a sample of 28 radio-loud AGNs \citep{Kang_2020}. We found that \ec could be ubiquitously (9 out 11) detected in radio AGNs with NuSTAR net counts above 10$^{4.5}$, and the ubiquitous detections of \ec in FR II galaxies indicate their X-ray emission is dominated by the thermal corona, instead of the jet. While for sources with lower NuSTAR counts, only a minor fraction of \ec detections (4 out of 17) were achieved. This motivates this work to perform systematic analyses of NuSTAR spectra of a sample of radio-quiet AGNs with sufficiently high signal to noise ratio of \Nu spectra (to avoid too many lower limits), and to statistically study the distribution of $E_{\rm cut}$/$kT_{\rm e}$, its dependence on other parameters, and the comparison with radio-loud AGNs.

\par The paper is organized as follows.  In \S \ref{sec:sample}, we present the sample selection and data reduction. The spectral fitting process as well as the fitting results are shown in \S \ref{sec:fitting}. Discussions are put in \S \ref{sec:dis}.

\section{The Sample and Data Reduction}\label{sec:sample}
\par We match the 817 Seyfert galaxies in the 105-month BAT catalogue \citep{Oh_2018} with the archival \Nu observations (as of October 2020). 
We drop observations with exposure time $<$ 3 ks,
or with total net counts (FPMA + FPMB)  $<$ 3000, for which no valid \ec measurement can be obtained.
We exclude a few exposures contaminated by solar activity or other unknown issues (through visually checking the images).
Furthermore, we exclude Compton-thick or heavily obscured sources (with  n$_{\rm H} > 10^{23} cm^{-2}$ fitted with a simple neutral absorber model). Based on the spectral fitting introduced in \S\ref{sec:fitting}, several observations with extremely hard spectra (photon index $\Gamma <$ 1.3) or poor fitting statistics ($\chi^2_\nu >$ 1.2), for which more complicated spectral models would be required, are also dropped. 
After these steps, 198 sources are kept, including 20 radio-loud sources and 178 radio-quiet sources. 

\par \citet{Kang_2020} presented a radio-loud sample of 28 sources with NuSTAR exposures, 20 of which are included in the sample described above, while the rest 8 sources are classified as ``beamed AGN'' in the BAT catalog \citep{Oh_2018}. Among them, 3C 279 is later found to be a jet-dominated blazar \citep[e.g.,][]{Blinov_2021} and is excluded from this work. Besides, we drop NGC 1275 (3C 84) due to the strong contamination from the diffuse thermal emission of the Perseus cluster to its spectra \citep{Rani_2018_1275}. 

\par For sources with multiple \Nu exposures observations, the ones with the most 3--78 keV net counts are adopted. 
Raw data are reduced using the NuSTAR Data Analysis Software within the latest version of HEASoft package (version 6.28), with calibration files CALDB version 20201101. These new versions of HEASoft and CALDB are applied to revise the recently noticed low-energy effective area issue of FPMA \citep{Madsen_2020}, which may partly account for the different fitting results from previous literature. The standard pipeline \textit{nupipeline} is used to generate the calibrated and cleaned event files. Following \citet{Kang_2020, Kang_2021}, each source spectrum is extracted in a circular region with a radius of 60\arcsec\ centered on each source using \textit{nuproduct}, while the background spectrum is derived using NUSKYBGD \citep{Wik_2014}, handling the spatially non-uniform background. As the last step, spectra are rebinned using \textit{grppha} to achieve a minimum of 50 counts bin$^{-1}$.

\begin{figure}
\centering
\subfloat{\includegraphics[width=0.47\textwidth]{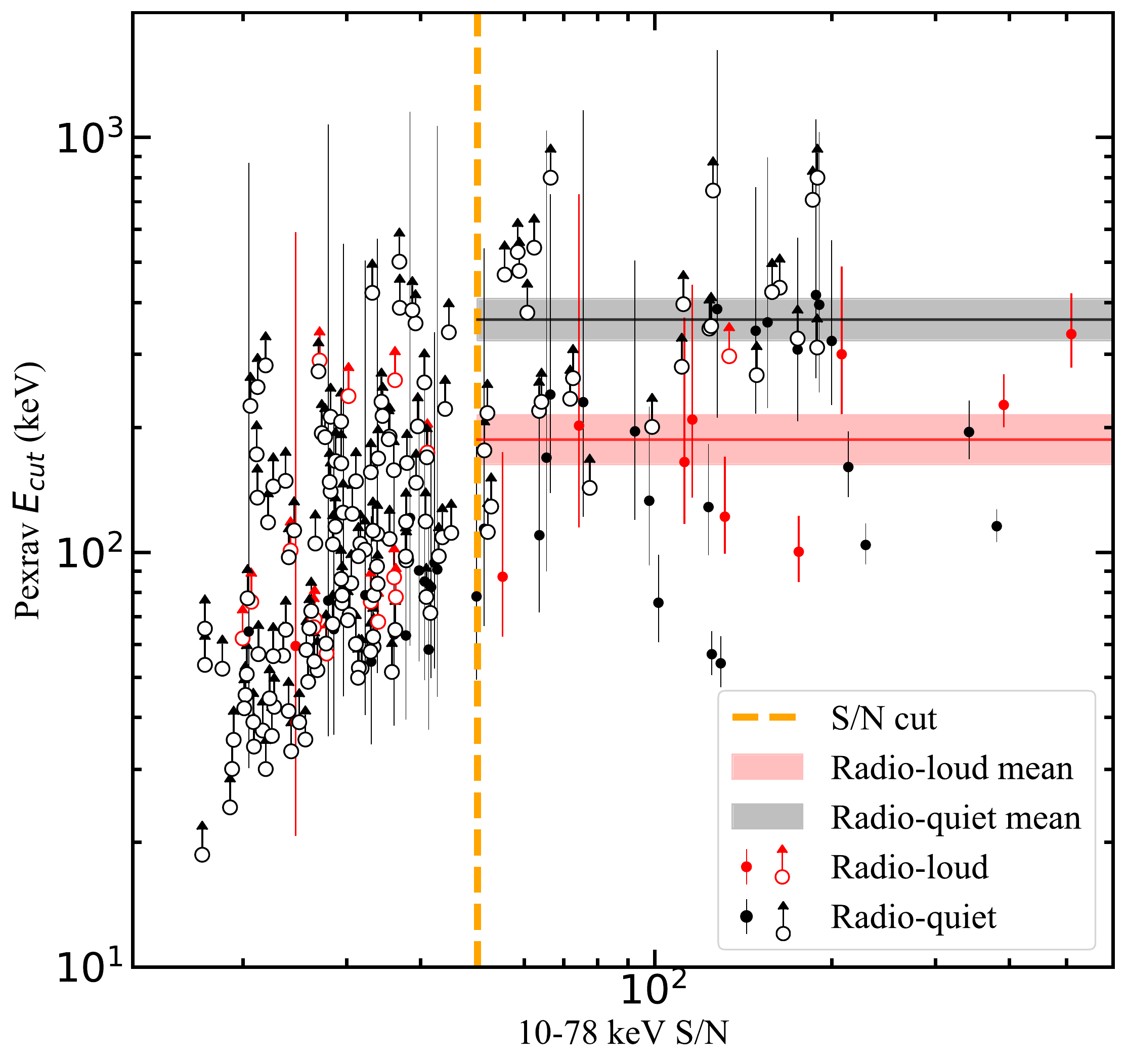}}
\caption{$E_{\rm cut}$ or lower limits from model $pexrav$ vs. 10--78 keV \Nu (FPMA) spectral S/N. A cut at 50 is adopted, sources below which are dropped. Mean values for radio-quiet and radio-loud samples are calculated using Kaplan-Meier estimator within ASURV in logarithm space (hereafter the same), and the shaded regions plot the 1$\sigma$ scatter of the mean derived through bootstrapping the corresponding sample (hereafter the same).}
\label{fig:SNcut}
\end{figure}  

\par We note the \ec measurement is profoundly affected by the quality of the spectra, particularly at the high energy band. In Fig. \ref{fig:SNcut} we plot the best-fit $E_{\rm cut}$ (or lower limits, derived through fitting \Nu spectra with $pexrav$, see \S\ref{sec:fitting}) for the 178 radio-quiet and 26 radio-loud AGNs, versus 10--78 keV S/N of \Nu FPMA net counts. 
Clearly the measurements of \ec for sources with low 10--78 keV S/N are dominated by poorly constrained lower limits for both radio-loud and radio-quiet sources. 
The lower limits systematically and significantly increase with 10--78 keV S/N at S/N $<$ 50 and the increase saturates at S/N $>$ 50. This indicates a threshold in S/N is essential to derive effective constraints to $E_{\rm cut}$. 
Thus in this work we focus only on sources with 
10--78 keV \Nu spectral S/N $>$ 50, including 50 radio-quiet and 10 radio-loud\footnote{Including 7 FR \uppercase\expandafter{\romannumeral2}, 2 FR \uppercase\expandafter{\romannumeral1}, and 1 core-dominated sources} sources (see Tab. \ref{tab:source_details}). 

\par We notice some \Nu observations have joint exposures from other missions like XMM-Newton or Swift. Those data are not included in this work mainly because different photon indices have been found between the spectra of \Nu and other missions \citep[e.g., ][]{gamma_xmm_1,gamma_xmm_2,gamma_xmm_3}, which may lead to significantly biased $E_{\rm cut}$/$kT_{\rm e}$ measurements. Such discrepancy is likely caused by the imperfect inter-instrument calibration, while the fact that joint exposures are not completely simultaneous (different start/end time, different livetime distribution) can also play a part due to rapid spectral variations.
Significant loss of the valuable NuSTAR exposure time would be unavoidable if we require perfect simultaneity between NuSTAR and exposures from other missions.
Considering the $E_{\rm cut}$/$kT_{\rm e}$ measurement is sensitive to the photon index, and to avoid the potential bias due to the fact that only a fraction of the exposures have quasi-simultaneous observations from other various missions, here we perform uniform spectral fitting to \Nu spectra alone for the whole sample. 

\startlongtable
\begin{deluxetable*}{cccccccc}
\tabletypesize{\scriptsize}
\renewcommand\tabcolsep{2.4pt}
\tablecaption {Sample Details\label{tab:source_details}}
\tablehead{
\colhead{Source} &\colhead{obsID} & \colhead{10--78 keV S/N}  &\colhead{Log M} & \colhead{ Log Lbol$_{ \rm 14-195 keV}$} & \colhead{Log $\lambda_{\rm edd}$} & \colhead{Log L$_{ \rm 0.1-200 keV}$} & \colhead{Compactness $l$} \\
\colhead{} & \colhead{} & \colhead{} & \colhead{$M_{\odot}$}   &  \colhead{erg/s} &  \colhead{}  &  \colhead{erg/s}  &  \colhead{} }  
\startdata
\multicolumn{8}{c}{Radio-quiet}\\
\hline
Mrk 1148 & 60160028002 & 51 & 7.82 & 45.3 & -0.64 & 44.8 & 161 \\
Fairall 9 & 60001130003 & 111 & 8.30 & 45.3 & -1.14 & 44.6 & 34 \\
NGC 931 & 60101002002 & 111 & 7.29 & 44.5 & -0.96 & 43.8 & 59 \\
HB89 0241+622 & 60160125002 & 66 & 8.09 & 45.6 & -0.60 & 44.6 & 62 \\
NGC 1566 & 80301601002 & 162 & 5.74 & 42.5 & -1.38 & 43.1 & 449 \\
1H 0419-577 & 60101039002 & 129 & 8.07 & 45.7 & -0.46 & 45.1 & 181 \\
Ark 120 & 60001044004 & 125 & 8.07 & 45.1 & -1.07 & 44.5 & 46 \\
ESO 362-18 & 60201046002 & 97 & 7.42 & 44.1 & -1.41 & 43.2 & 11 \\
2MASX J05210136-2521450 & 60201022002 & 52 & - & - & - & 43.8 & - \\
NGC 2110 & 60061061002 & 174 & 9.25 & 44.6 & -2.79 & 44.2 & 1.5 \\
MCG +08-11-011 & 60201027002 & 187 & 7.62 & 45.0 & -0.76 & 44.3 & 78 \\
MCG +04-22-042 & 60061092002 & 51 & 7.34 & 44.9 & -0.56 & 44.2 & 147 \\
Mrk 110 & 60201025002 & 213 & 7.29 & 45.1 & -0.29 & 44.6 & 354 \\
NGC 2992 & 90501623002 & 190 & 5.42 & 43.4 & -0.14 & 43.7 & 3413 \\
MCG-05-23-016 & 60001046008 & 380 & 5.86 & 44.4 & 0.40 & 43.7 & 1172 \\
NGC 3227 & 60202002014 & 148 & 6.77 & 43.6 & -1.33 & 42.9 & 22 \\
NGC 3516 & 60002042004 & 58 & 7.39 & 44.5 & -1.03 & 42.7 & 4.2 \\
HE 1136-2304 & 80002031003 & 65 & 7.62 & 44.4 & -1.40 & 43.8 & 26 \\
NGC 3783 & 60101110002 & 123 & 7.37 & 44.6 & -0.90 & 43.4 & 20 \\
UGC 06728 & 60376007002 & 75 & 5.66 & 43.3 & -0.51 & 44.7 & 21819 \\
2MASX J11454045-1827149 & 60302002006 & 63 & 7.31 & 45.0 & -0.42 & 44.3 & 181 \\
NGC 3998 & 60201050002 & 63 & 8.93 & 42.8 & -4.23 & 41.8 & 0.01 \\
NGC 4051 & 60401009002 & 188 & 6.13 & 42.9 & -1.34 & 41.8 & 8.2 \\
Mrk 766 & 60001048002 & 98 & 6.82 & 43.8 & -1.15 & 43.3 & 53 \\
NGC 4593 & 60001149008 & 66 & 6.88 & 44.0 & -1.05 & 43.2 & 38 \\
WKK 1263 & 60160510002 & 58 & 8.25 & 44.7 & -1.66 & 44.2 & 17 \\
MCG-06-30-015 & 60001047003 & 185 & 5.82 & 43.8 & -0.11 & 43.2 & 444 \\
NGC 5273 & 60061350002 & 55 & 6.66 & 42.5 & -2.26 & 42.3 & 7.9 \\
4U 1344-60 & 60201041002 & 174 & 7.32 & 44.5 & -0.95 & 43.7 & 48 \\
IC 4329A & 60001045002 & 341 & 7.84 & 45.1 & -0.85 & 44.3 & 59 \\
Mrk 279 & 60160562002 & 62 & 7.43 & 44.8 & -0.75 & 44.2 & 119 \\
NGC 5506 & 60061323002 & 158 & 5.62 & 44.1 & 0.37 & 43.3 & 794 \\
NGC 5548 & 60002044006 & 123 & 7.72 & 44.6 & -1.24 & 44.0 & 31 \\
WKK 4438 & 60401022002 & 71 & 6.86 & 44.0 & -0.98 & 43.2 & 39 \\
Mrk 841 & 60101023002 & 51 & 7.81 & 45.0 & -0.99 & 44.3 & 55 \\
AX J1737.4-2907 & 60301010002 & 101 & - & - & - & 44.2 & - \\
2MASXi J1802473-145454 & 60160680002 & 52 & 7.76 & 45.0 & -0.92 & 44.4 & 79 \\
ESO 141-G 055 & 60201042002 & 124 & 8.07 & 45.1 & -1.06 & 44.4 & 39 \\
2MASX J19373299-0613046 & 60101003002 & 77 & 6.56 & 43.6 & -1.04 & 43.1 & 57 \\
NGC 6814 & 60201028002 & 188 & 7.04 & 43.6 & -1.58 & 42.9 & 12 \\
Mrk 509 & 60101043002 & 228 & 8.05 & 45.3 & -0.86 & 44.6 & 63 \\
SWIFT J212745.6+565636 & 60402008004 & 124 & 7.20 (1) & - & - & 43.7 & 51 \\
NGC 7172 & 60061308002 & 127 & 8.45 & 44.3 & -2.31 & 43.6 & 2.8 \\
NGC 7314 & 60201031002 & 148 & 4.99 & 43.2 & 0.12 & 42.7 & 970 \\
Mrk 915 & 60002060002 & 60 & 7.71 & 44.5 & -1.33 & 43.7 & 18 \\
MR 2251-178 & 60102025004 & 92 & 8.44 & 45.9 & -0.66 & 45.3 & 126 \\
NGC 7469 & 60101001014 & 72 & 6.96 & 44.5 & -0.60 & 41.8 & 1.4 \\
Mrk 926 & 60201029002 & 199 & 8.55 & 45.7 & -1.01 & 45.0 & 56 \\
NGC 4579 & 60201051002 & 64 & 7.80 & - & - & 42.2 & 0.43 \\
M 81 & 60101049002 & 155 & 7.90 & 41.3 & -4.72 & 41.1 & 0.03 \\
\hline
\multicolumn{8}{c}{Radio-loud}\\
\hline
3C 109 & 60301011004 & 55 & 8.30 (2) & 47.4 & 0.98 & 45.8 & 539 \\
3C 111 & 60202061004 & 112 & 8.27 & 45.7 & -0.67 & 44.9 & 82 \\
3C 120 & 60001042003 & 207 & 7.74 & 45.3 & -0.59 & 44.7 & 152 \\
PicA & 60101047002 & 74 & 7.60 (3)& 44.9 & -0.79 & 43.9 & 38 \\
3C 273 & 10002020001 & 391 & 8.84 & 47.4 & 0.41 & 46.4 & 641 \\
CentaurusA & 60001081002 & 509 & 7.74 (4) & 43.3 & -2.61 & 42.7 & 1.6 \\
3C 382 & 60001084002 & 133 & 8.19 & 45.7 & -0.58 & 45.0 & 116 \\
3C 390.3 & 60001082003 & 115 & 8.64 & 45.8 & -0.99 & 45.1 & 50 \\
4C 74.26 & 60001080006 & 131 & 9.60 & 46.1 & -1.67 & 45.4 & 12 \\
IGR J21247+5058 & 60301005002 & 175 & 7.63 & 44.9 & -0.85 & 44.5 & 145 \\
\enddata 
\tablecomments{Sources are ordered by BAT ID. The 10--78 keV signal-to-noise ratios are calculated using FPMA spectra. The blackhole masses are from \citet{Koss_2017}, unless marked with a number referring to the following literature.(1) \citet{Malizia_2008}; (2) \citet{3C109_mass}; (3) \citet{PicA_4579_mass}; (4) \citet{CenA_mass} . Lbol$_{ \rm 14-195 keV}$ is the bolometric luminosity estimated by the BAT 14--195 keV flux \citep{Koss_2017} and used for $\lambda_{\rm edd}$ calculation. 
L$_{ \rm 0.1-200 keV}$ is the unabsorbed 0.1--200 keV luminosity, extrapolated using the best-fit results of $pexrav$ to NuSTAR spectra and adopting the redshifts from the 105-month BAT catalogue and $H_{0} = 70$ km s$^{-1}$ Mpc$^{-1}$. The compactness parameter $l$ is derived from L$_{ \rm 0.1-200 keV}$. }
\end{deluxetable*}

\section{Spectral Fitting}\label{sec:fitting}

\par Spectral fitting is carried out within the 3--78 keV band using XSPEC \citep{Arnaud_1996}. $\chi ^2$ statistics is adopted and all the errors together with the upper/lower limits in this paper correspond to 90\% confidence level with $\Delta \chi^2 = 2.71$, unless otherwise stated. The relative element abundance is set to the default in XSPEC, given by \cite{ANDERS1989197}. For each observation, the spectra of FPMA and FPMB are jointly fitted with a cross-normalization \citep{Madsen_2015}.

\par In this paper we intend to perform uniform measurements of the \ec / $T_{\rm e}$ for the radio-quiet and radio-loud samples and bring them into comparison. In order to guarantee such comparison is model-independent, various models are employed, 
including 
$pexrav$, $relxill$, and $relxillcp$. 

\par $pexrav$ \citep{Magdziarz_1995} is the model we used to fit the radio-loud sample in \citet{Kang_2020}, which fits the spectra with an exponentially cutoff power law plus a neutral reflection component, and is the most widely used model in \ec measurement \citep[e.g.,][]{Molina_2019, Rani_2019, Panagiotou_2020, Balokovi_2020, Kang_2021}. For simplicity, the solar element abundance for the reflector and an inclination of cos$i$ = 0.45 are adopted, which are the default values of the model. We allow the photon index $\Gamma$, \ec and the reflection scaling factor $R$ free to vary.

\par  $relxill$ \citep{Garc_2014} also models the underlying continuum with a cutoff powerlaw, but convolves the reflection component with disc relativistic broadening effect. However, some parameters are hard to constrain even with these high-quality \Nu spectra and hence have to be frozen. The inner and outer radius of the accretion disk, $Rin$ and $Rout$, are fixed at 1 ISCO and 400 gravitational radii respectively as the default of the model. Besides we fix the blackhole spin $a = 0.998$\footnote{The fitting results however are insensitive to this choice.} and the inclination angle $i = 30^{\circ}$. The accretion disk is presumed to be neutral and have the solar iron abundance, with the corresponding parameter $logxi$ and $Afe$ fixed at 0 and 1, respectively. We assume a disk with constant emissivity, setting the emissivity parameter $Index2$ tied with $Index1$. The free parameters include $Index1$, $\Gamma$, $E_{\rm cut}$ and the reflection fraction (with different definition from the $R$ in $pexrav$). 

\par A Comptonization model, $relxillcp$, is also adopted to directly measure the coronal temperature $T_{\rm e}$. $relxillcp$ is a Comptonization version of $relxill$, replacing the cutoff power law with a nthcomp continuum. Other parameters are set in the same way as $relxill$.

\par Meanwhile, a common component $zphabs$ is added to all three models to represent the intrinsic photoelectric absorption, with the Galactic absorption ignored due to its inappreciable influence on \Nu spectra. As for the Fe K$\alpha$ lines, in $relxill$ and $relxillcp$ the continuum reflection component and the Fe K$\alpha$ line are jointly fitted, while a $zgauss$ is added to $pexrav$ to describe the Fe K$\alpha$ line. Since a relativistically broadened Fe K$\alpha$ line can not be well constrained in the majority of observations, we deal with the Gaussian component as follows. In the first place we fix the line at 6.4 keV in the rest frame and the line width at 19 eV \citep[the mean Fe K$\alpha$ line width in AGNs measured with Chandra HETG,][]{Shu_2010} to model a neutral narrow Fe K$\alpha$ line. Then we allow the line width free to vary. If a variable line width prominently improves the fitting ($\Delta \chi^2 > 5$), the corresponding fitting results are adopted.

\par We summarize below the three models adopted in the XSPEC term and the corresponding free parameters. 
\begin{itemize}
\item{ 
$zphabs*(pexrav + zgauss)$\\
Free parameters include absorption column density $n_{\rm H}$, photon index $\Gamma$, high energy cutoff \ec and the strength of the reflection component R.
}
\item{
$zphabs*relxill$\\
$n_{\rm H}$, $\Gamma$, $E_{\rm cut}$, emissivity parameter $Index1$ and the reflection fraction. 
 }
 \item{
$zphabs*relxillcp$\\
Same as $relxill$, except that \ec is replaced with $T_{\rm e}$. 
 }
\end{itemize}
The best-fitting results of the key parameters are shown in Tab. \ref{tab:fitting_result}. 
In a few sources the spectral fitting yields very high lower limits of $E_{\rm cut}$, up to 2360 keV (see \S4 for further discussion on reliability of such high lower limits of $E_{\rm cut}$).
For the two sources with \ec lower limits above 800 keV ($pexrav$ results; NGC 4051, $>$ 2360 keV; NGC 4593, $>$ 1420 keV), we manually and conservatively set their \ec lower limits at 800 keV. Simply adopting their best-fit lower limits would further strengthen the results of this work.

\startlongtable
\begin{deluxetable*}{cccccccccc}
\tabletypesize{\scriptsize}
\renewcommand\tabcolsep{2.4pt}
\tablecaption{Spectral Fitting Results \label{tab:fitting_result}}
\tablehead{
\colhead{Source} &\colhead{obsID} & \colhead{$\Gamma^{\rm pexrav}$} & \colhead{$R^{\rm pexrav}$}  &  \colhead{$E_{\rm cut}^{\rm pexrav}$}  &  \colhead{$\chi_{\rm pexrav}^2/dof$}  & \colhead{$E_{\rm cut}^{\rm relxill}$} &  \colhead{$\chi_{\rm relxill}^2/dof$} &\colhead{$T_{\rm e}^{\rm relxillcp}$} &  \colhead{$\chi_{\rm relxillcp}^2/dof$}\\
\colhead{} &  \colhead{}  &  \colhead{}  &  \colhead{} &\colhead{keV}  &  \colhead{} & \colhead{keV} &  \colhead{} &\colhead{keV} &  \colhead{} }  
\startdata
\multicolumn{10}{c}{Radio-quiet}\\
\hline
Mrk 1148 & 60160028002 & $1.79_{-0.08}^{+0.13}$ & $ < 0.45 $ & $113_{-47}^{+427}$ &  0.91 & $ > 65 $ &  0.92 & $ > 18 $ & 0.92 \\
Fairall 9 & 60001130003 & $1.96_{-0.03}^{+0.06}$ & $0.71_{-0.17}^{+0.22}$ & $ > 396 $ &  0.91 & $ > 400 $ &  0.94 & $ > 182 $ & 0.96 \\
NGC 931 & 60101002002 & $1.88_{-0.06}^{+0.06}$ & $0.70_{-0.19}^{+0.21}$ & $ > 280 $ &  0.85 & $ > 293 $ &  0.86 & $ > 138 $ & 0.86 \\
HB89 0241+622 & 60160125002 & $1.70_{-0.06}^{+0.06}$ & $0.73_{-0.27}^{+0.33}$ & $240_{-101}^{+489}$ &  0.97 & $ > 158 $ &  1.02 & $ > 50 $ & 1.02 \\
NGC 1566 & 80301601002 & $1.84_{-0.04}^{+0.05}$ & $0.80_{-0.14}^{+0.15}$ & $ > 434 $ &  0.93 & $ > 511 $ &  0.93 & $ > 173 $ & 0.93 \\
1H 0419-577 & 60101039002 & $1.64_{-0.05}^{+0.06}$ & $0.38_{-0.13}^{+0.15}$ & $54_{-6}^{+8}$ &  0.99 & $54_{-6}^{+8}$ &  0.99 & $16_{-1}^{+1}$  & 1.00 \\
Ark 120 & 60001044004 & $1.98_{-0.03}^{+0.03}$ & $0.58_{-0.12}^{+0.14}$ & $ > 744 $ &  1.06 & $ > 414 $ &  1.10 & $ > 213 $ & 1.12 \\
ESO 362-18 & 60201046002 & $1.57_{-0.08}^{+0.09}$ & $0.58_{-0.22}^{+0.26}$ & $133_{-40}^{+91}$ &  1.03 & $135_{-33}^{+92}$ &  1.08 & $ > 34 $ & 1.09 \\
2MASX J05210136-2521450 & 60201022002 & $2.06_{-0.12}^{+0.15}$ & $0.33_{-0.30}^{+0.74}$ & $ > 111 $ &  0.99 & $ > 129 $ &  1.00 & $ > 49 $ & 1.00 \\
NGC 2110 & 60061061002 & $1.67_{-0.03}^{+0.03}$ & $ < 0.03 $ & $ > 327 $ &  0.95 & $ > 382 $ &  0.96 & $ > 217 $ & 0.99 \\
MCG +08-11-011 & 60201027002 & $1.81_{-0.02}^{+0.04}$ & $0.26_{-0.09}^{+0.10}$ & $417_{-154}^{+688}$ &  1.03 & $ > 302 $ &  1.09 & $ > 244 $ & 1.10 \\
MCG +04-22-042 & 60061092002 & $1.95_{-0.09}^{+0.10}$ & $0.59_{-0.33}^{+0.44}$ & $ > 216 $ &  0.87 & $ > 167 $ &  0.88 & $ > 37 $ & 0.88 \\
Mrk 110 & 60201025002 & $1.74_{-0.01}^{+0.01}$ & $ < 0.04 $ & $160_{-24}^{+35}$ &  1.05 & $159_{-32}^{+43}$ &  1.10 & $57_{-18}^{+54}$  & 1.11 \\
NGC 2992 & 90501623002 & $1.68_{-0.04}^{+0.04}$ & $0.08_{-0.07}^{+0.08}$ & $395_{-152}^{+636}$ &  1.05 & $ > 316 $ &  1.15 & $ > 260 $ & 1.16 \\
MCG -05-23-016 & 60001046008 & $1.72_{-0.02}^{+0.02}$ & $0.45_{-0.05}^{+0.05}$ & $115_{-9}^{+11}$ &  1.10 & $125_{-8}^{+10}$ &  1.19 & $41_{-3}^{+3}$  & 1.24 \\
NGC 3227 & 60202002014 & $1.90_{-0.05}^{+0.05}$ & $1.21_{-0.19}^{+0.22}$ & $342_{-125}^{+417}$ &  1.01 & $ > 251 $ &  1.01 & $ > 83 $ & 1.01 \\
NGC 3516 & 60002042004 & $1.68_{-0.09}^{+0.09}$ & $0.65_{-0.30}^{+0.39}$ & $ > 476 $ &  1.10 & $ > 368 $ &  1.17 & $ > 114 $ & 1.18 \\
HE 1136-2304 & 80002031003 & $1.69_{-0.10}^{+0.10}$ & $ < 0.48 $ & $169_{-79}^{+871}$ &  1.00 & $160_{-71}^{+573}$ &  1.00 & $ > 21 $ & 1.01 \\
NGC 3783 & 60101110002 & $1.94_{-0.07}^{+0.07}$ & $1.58_{-0.28}^{+0.33}$ & $ > 346 $ &  1.05 & $ > 432 $ &  1.04 & $ > 150 $ & 1.05 \\
UGC 06728 & 60376007002 & $1.80_{-0.10}^{+0.10}$ & $0.75_{-0.28}^{+0.33}$ & $230_{-108}^{+933}$ &  1.01 & $183_{-62}^{+452}$ &  1.01 & $ > 26 $ & 1.01 \\
2MASX J11454045-1827149 & 60302002006 & $1.79_{-0.08}^{+0.11}$ & $0.43_{-0.27}^{+0.33}$ & $109_{-38}^{+124}$ &  0.88 & $105_{-33}^{+98}$ &  0.88 & $29_{-10}^{+44}$  & 0.89 \\
NGC 3998 & 60201050002 & $1.96_{-0.07}^{+0.08}$ & $ < 0.34 $ & $ > 219 $ &  0.96 & $ > 201 $ &  0.97 & $ > 47 $ & 0.97 \\
NGC 4051 & 60401009002 & $2.05_{-0.03}^{+0.03}$ & $2.04_{-0.20}^{+0.22}$ & $ > 800 $ &  1.04 & $ > 800 $ &  1.02 & $ > 270 $ & 1.05 \\
Mrk 766 & 60001048002 & $2.30_{-0.07}^{+0.07}$ & $1.76_{-0.34}^{+0.40}$ & $ > 200 $ &  1.05 & $ > 352 $ &  1.05 & $ > 157 $ & 1.06 \\
NGC 4593 & 60001149008 & $1.83_{-0.05}^{+0.05}$ & $0.63_{-0.21}^{+0.25}$ & $ > 800 $ &  0.99 & $ > 577 $ &  1.00 & $ > 144 $ & 1.02 \\
WKK 1263 & 60160510002 & $1.79_{-0.09}^{+0.09}$ & $ < 0.50 $ & $ > 529 $ &  0.86 & $ > 374 $ &  0.86 & $ > 72 $ & 0.87 \\
MCG -06-30-015 & 60001047003 & $2.29_{-0.04}^{+0.02}$ & $1.83_{-0.20}^{+0.21}$ & $ > 707 $ &  1.07 & $ > 720 $ &  1.05 & $ > 280 $ & 1.07 \\
NGC 5273 & 60061350002 & $1.90_{-0.11}^{+0.11}$ & $1.30_{-0.50}^{+0.66}$ & $ > 467 $ &  1.11 & $ > 362 $ &  1.11 & $ > 85 $ & 1.12 \\
4U 1344-60 & 60201041002 & $1.90_{-0.05}^{+0.05}$ & $0.92_{-0.15}^{+0.17}$ & $308_{-101}^{+265}$ &  1.11 & $337_{-112}^{+204}$ &  1.12 & $ > 104 $ & 1.13 \\
IC 4329A & 60001045002 & $1.72_{-0.02}^{+0.02}$ & $0.32_{-0.05}^{+0.05}$ & $195_{-27}^{+37}$ &  1.03 & $215_{-33}^{+37}$ &  1.07 & $71_{-15}^{+37}$  & 1.09 \\
Mrk 279 & 60160562002 & $1.90_{-0.04}^{+0.05}$ & $0.19_{-0.17}^{+0.20}$ & $ > 542 $ &  1.01 & $ > 231 $ &  1.07 & $ > 84 $ & 1.07 \\
NGC 5506 & 60061323002 & $1.90_{-0.06}^{+0.05}$ & $1.29_{-0.19}^{+0.22}$ & $ > 424 $ &  1.08 & $ > 551 $ &  1.06 & $ > 211 $ & 1.07 \\
NGC 5548 & 60002044006 & $1.69_{-0.06}^{+0.06}$ & $0.62_{-0.17}^{+0.19}$ & $128_{-30}^{+53}$ &  0.99 & $126_{-28}^{+45}$ &  1.00 & $36_{-8}^{+11}$  & 1.01 \\
WKK 4438 & 60401022002 & $2.00_{-0.05}^{+0.08}$ & $1.11_{-0.33}^{+0.44}$ & $ > 234 $ &  0.92 & $ > 263 $ &  0.93 & $ > 81 $ & 0.94 \\
Mrk 841 & 60101023002 & $1.89_{-0.12}^{+0.06}$ & $0.45_{-0.33}^{+0.45}$ & $ > 176 $ &  1.02 & $ > 154 $ &  1.02 & $ > 44 $ & 1.02 \\
AX J1737.4-2907 & 60301010002 & $1.79_{-0.08}^{+0.08}$ & $0.94_{-0.25}^{+0.29}$ & $75_{-14}^{+23}$ &  1.05 & $112_{-36}^{+95}$ &  1.04 & $35_{-14}^{+184}$  & 1.04 \\
2MASXi J1802473-145454 & 60160680002 & $1.76_{-0.08}^{+0.09}$ & $ < 0.41 $ & $ > 128 $ &  1.03 & $ > 135 $ &  1.08 & $ > 53 $ & 1.08 \\
ESO 141- G 055 & 60201042002 & $1.92_{-0.03}^{+0.03}$ & $0.67_{-0.15}^{+0.17}$ & $ > 351 $ &  1.04 & $ > 293 $ &  1.05 & $ > 125 $ & 1.05 \\
2MASX J19373299-0613046 & 60101003002 & $2.45_{-0.11}^{+0.16}$ & $2.01_{-0.65}^{+1.39}$ & $ > 143 $ &  0.99 & $ > 217 $ &  1.13 & $ > 137 $ & 1.14 \\
NGC 6814 & 60201028002 & $1.83_{-0.03}^{+0.04}$ & $0.46_{-0.10}^{+0.11}$ & $ > 311 $ &  1.06 & $371_{-108}^{+318}$ &  1.10 & $ > 147 $ & 1.11 \\
Mrk 509 & 60101043002 & $1.75_{-0.02}^{+0.02}$ & $0.41_{-0.07}^{+0.08}$ & $104_{-10}^{+13}$ &  1.06 & $98_{-8}^{+13}$ &  1.08 & $24_{-2}^{+2}$  & 1.12 \\
SWIFT J212745.6+565636 & 60402008004 & $2.10_{-0.04}^{+0.05}$ & $1.77_{-0.33}^{+0.45}$ & $56_{-6}^{+7}$ &  1.06 & $63_{-8}^{+11}$ &  1.07 & $21_{-3}^{+8}$  & 1.07 \\
NGC 7172 & 60061308002 & $1.84_{-0.06}^{+0.06}$ & $0.68_{-0.17}^{+0.19}$ & $385_{-174}^{+1239}$ &  1.05 & $337_{-137}^{+523}$ &  1.05 & $ > 54 $ & 1.05 \\
NGC 7314 & 60201031002 & $2.06_{-0.05}^{+0.05}$ & $1.18_{-0.19}^{+0.21}$ & $ > 267 $ &  1.05 & $ > 346 $ &  1.07 & $ > 188 $ & 1.07 \\
Mrk 915 & 60002060002 & $1.81_{-0.09}^{+0.10}$ & $0.37_{-0.27}^{+0.34}$ & $ > 378 $ &  1.03 & $ > 286 $ &  1.05 & $ > 69 $ & 1.06 \\
MR 2251-178 & 60102025004 & $1.77_{-0.07}^{+0.07}$ & $ < 0.25 $ & $195_{-75}^{+310}$ &  1.02 & $ > 117 $ &  1.01 & $37_{-12}^{+97}$  & 1.01 \\
NGC 7469 & 60101001014 & $1.85_{-0.05}^{+0.08}$ & $0.41_{-0.21}^{+0.30}$ & $ > 262 $ &  0.88 & $ > 242 $ &  0.89 & $ > 77 $ & 0.89 \\
Mrk 926 & 60201029002 & $1.73_{-0.02}^{+0.02}$ & $ < 0.10 $ & $323_{-96}^{+241}$ &  1.05 & $292_{-87}^{+178}$ &  1.11 & $ > 83 $ & 1.11 \\
NGC 4579 & 60201051002 & $1.88_{-0.08}^{+0.04}$ & $ < 0.15 $ & $ > 230 $ &  1.02 & $ > 93 $ &  1.08 & $ > 49 $ & 1.08 \\
M 81 & 60101049002 & $1.88_{-0.02}^{+0.02}$ & $ < 0.05 $ & $358_{-135}^{+538}$ &  1.00 & $225_{-85}^{+233}$ &  1.10 & $ > 83 $ & 1.11 \\
\hline
\multicolumn{10}{c}{Radio-loud}\\
\hline
3C 109 & 60301011004 & $1.64_{-0.08}^{+0.16}$ & $0.32_{-0.24}^{+0.32}$ & $87_{-24}^{+86}$ & 0.95 & $88_{-25}^{+97}$ & 0.96 & $30_{-11}^{+60}$& 0.97 \\
3C 111 & 60202061004 & $1.70_{-0.04}^{+0.06}$ & $ < 0.08 $ & $165_{-47}^{+202}$ & 1.07 & $174_{-57}^{+166}$ & 1.10 & $ > 35 $ & 1.11 \\
3C 120 & 60001042003 & $1.86_{-0.03}^{+0.03}$ & $0.40_{-0.08}^{+0.09}$ & $300_{-85}^{+188}$ & 1.01 & $289_{-80}^{+138}$ & 1.02 & $ > 91 $ & 1.02 \\
PicA & 60101047002 & $1.72_{-0.04}^{+0.04}$ & $ < 0.10 $ & $202_{-87}^{+527}$ & 0.98 & $161_{-74}^{+754}$ & 1.00 & $ > 29 $ & 1.01 \\
3C 273 & 10002020001 & $1.62_{-0.01}^{+0.02}$ & $0.05_{-0.03}^{+0.03}$ & $226_{-26}^{+42}$ & 1.02 & $ > 237 $ & 1.03 & $ > 79 $ & 1.03 \\
CentaurusA & 60001081002 & $1.75_{-0.01}^{+0.01}$ & $ < 0.01 $ & $335_{-56}^{+85}$ & 1.00 & $209_{-24}^{+34}$ & 1.04 & $101_{-24}^{+85}$& 1.08 \\
3C 382 & 60001084002 & $1.76_{-0.05}^{+0.04}$ & $ < 0.13 $ & $ > 297 $ & 0.95 & $ > 268 $ & 0.98 & $ > 105 $ & 0.98 \\
3C 390.3 & 60001082003 & $1.72_{-0.06}^{+0.06}$ & $0.14_{-0.12}^{+0.14}$ & $208_{-73}^{+232}$ & 0.98 & $235_{-85}^{+267}$ & 1.00 & $ > 46 $ & 1.00 \\
4C 74.26 & 60001080006 & $1.80_{-0.04}^{+0.07}$ & $0.66_{-0.15}^{+0.18}$ & $121_{-22}^{+48}$ & 0.99 & $165_{-42}^{+85}$ & 1.01 & $62_{-23}^{+265}$& 1.01 \\
IGR J21247+5058 & 60301005002 & $1.63_{-0.02}^{+0.04}$ & $ < 0.11 $ & $100_{-15}^{+22}$ & 1.07 & $102_{-16}^{+22}$ & 1.08 & $26_{-3}^{+6}$& 1.11 \\
\enddata 
\end{deluxetable*}

\section{Discussion}\label{sec:dis}

\par
\begin{figure}
\centering
\subfloat{\includegraphics[width=0.46\textwidth]{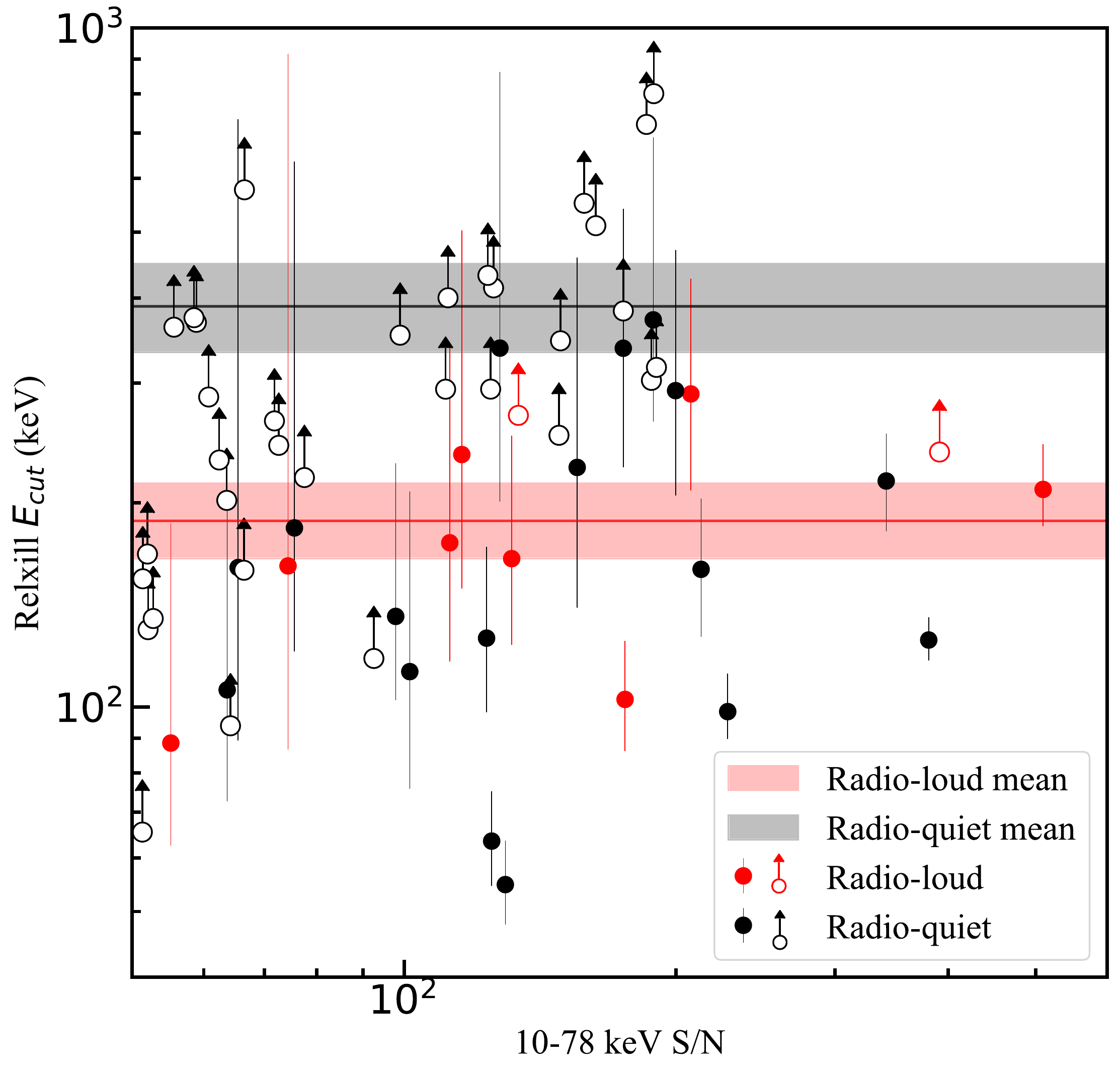}}\\
\subfloat{\includegraphics[width=0.46\textwidth]{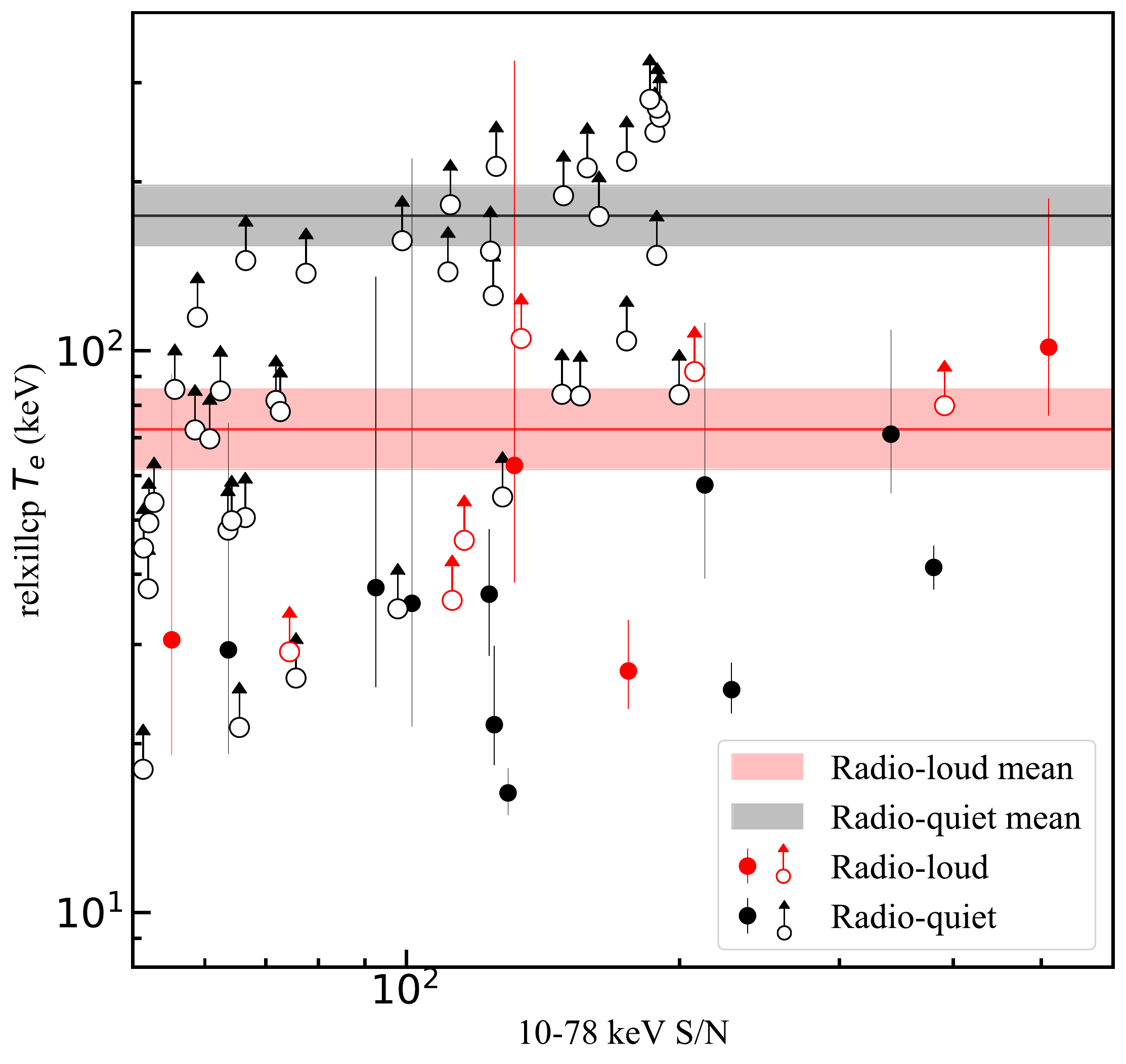}}
\caption{
Similar to Fig. \ref{fig:SNcut} but with $E_{\rm cut}$ (left) and $T_{\rm e}$ (right) derived from $relxill$ and $relxillcp$, respectively.
}
\label{fig:ETSN}
\end{figure}

 The best-fit $E_{\rm cut}$ from $pexrav$ is presented in Fig. \ref{fig:SNcut}.
We plot the $E_{\rm cut}$/$T_{\rm e}$ from the other two models versus 10--78 keV S/N in Fig. \ref{fig:ETSN}. Similar to \citet{Kang_2020}, we find the \ec in this radio-loud sample can be well constrained as long as the spectra have enough S/N. With $pexrav$ we obtain \ec measurements for 9 out 10 radio-loud sources with 10--78 keV S/N $>$ 50. The only radio-loud source without \ec detection is 3C 382, for which \ec detection was reported in another \Nu exposure with slightly less \Nu net counts than the one adopted in this work.

\begin{deluxetable}{cccc}
\tabletypesize{\scriptsize}
\tablecaption{The mean $E_{\rm cut}$/$T_{\rm e}$ of our RQ and RL samples. The last row presents the statistical significance of the difference in the mean value between two samples. \label{tab:Mean}}
\tablehead{
\colhead{} &  \colhead{$pexrav$ $E_{\rm cut}$}   & \colhead{$relxill$ $E_{\rm cut}$} & \colhead{$relxillcp$ $T_{\rm e}$}}
\startdata
RQ (keV) & $364^{+45}_{-40}$ & $390^{+60}_{-52}$ & $174^{+23}_{-20}$  \\
RL (keV) & $187^{+27}_{-24}$ & $188^{+26}_{-23}$ & $72^{+13}_{-11}$  \\
Significance ($\sigma$) & 3.6 & 3.4 & 4.2 \\
\enddata 
\end{deluxetable}

\par However, as shown in Fig. \ref{fig:SNcut} and Fig. \ref{fig:ETSN}, the case is markedly different in the radio-quiet sample where only lower limits to \ec could be obtained for 28 out of 50 sources ($pexrav$ results). 
In Fig. \ref{fig:SNcut} \& \ref{fig:ETSN} we also plot the mean $E_{\rm cut}$/$T_{\rm e}$ of the radio-quiet and loud samples. We adopt
the so-called survival statistics within the package ASURV \citep{Feigelson_1985} to take the lower limits into account. We employ the Kaplan-Meier estimator to estimate the mean of $E_{\rm cut}$/$T_{\rm e}$ for the two samples. As the Kaplan-Meier estimator is exceedingly sensitive to the value of the maximums, the calculation is performed in the logarithm space to weaken the imbalance of statistical weights\footnote{The derived mean is like the traditional geometric mean}. Since the dispersion given by the Kaplan-Meier estimator could be underestimated, we conservatively bootstrap the corresponding samples to 
obtain the dispersion to the mean. As shown in Tab. \ref{tab:Mean}, the mean of $E_{\rm cut}$/$T_{\rm e}$ of the radio-quiet sample is remarkably larger than that of the radio-loud one at a level above 3$\sigma$ for all three models.

\par We note 10 out of the 50 radio-quiet sources have considerably high \ec lower limits ($>$ 400 keV in $pexrav$ model), while all \ec measurements or lower limit from the radio-loud sample are below 400 keV. 
Note excluding these 10 large \ec lower limits would yield a lower average \ec for the radio-quiet sample (mean $pexrav$ \ec = 248$^
{+26}_{-24}$ keV), and the difference between radio-quiet and radio-loud samples is no longer statistically significant.
We therefore carefully further inspect these 10 individual sources in \S\ref{sec:literature} (ordered from low to high by the \ec lower limit) through comparing with results reported in literature. 

\subsection{Notes on sources with extraordinarily high $E_{cut}$ lower limits}\label{sec:literature}

\begin{enumerate}
\item NGC 5506, Sy 1.9, \ec $>$ 424 keV ($pexrav$, this work, hereafter the same for the rest 9 sources).
Consistently, 
\citet{Matt_2015} reported an \ec = $720_{-190}^{+130}$ keV and a 3$\sigma$ lower limit of 350 keV; \citet{Sun_2018} reported an \ec = $500_{-240}^{+100}$ keV; \citet{Panagiotou_2020} reported a 1$\sigma$ lower limit of 8400 keV. The only exception came from \citet{Balokovi_2020}, which reported an \ec = 110 $\pm$ 10 keV with contemporaneous Swift/BAT data. \citet{Balokovi_2020} claimed in its appendix that the BAT spectrum shows a much smaller cutoff than the \Nu one. \ec variation and background subtraction may have played a role.  

\item NGC 1566, Sy 1.5, \ec 
$>$ 434 keV. \citet{Akylas_2021} reported an \ec = $336_{-140}^{+646}$ keV, while \citet{Parker_2019} reported an incompatible result of \ec = $167 \pm 3$ keV. A possible reason is that \citet{Parker_2019} used quasi-simultaneous XMM-Newton data, which has a photon index $\Gamma_{\rm pn} \sim 1.43$, quite different from our result ($\Gamma \sim 1.84$). Besides, the reflection fraction $R_{relxill}$ is $\sim$ 0.09, smaller than that in this work ($R_{relxill}$ $\sim$ 0.25). The data are actually barely simultaneous, considering a start time offset of $\sim$ 10 ks and the fact that \Nu exposure is 57 ks while PN exposure is 100 ks. Meanwhile, the inter-instrument calibration issue and the pile-up effect in PN data may also have played a part here.

\item NGC 5273, Sy 1.5, \ec $>$ 467 keV. \citet{Panagiotou_2020} reported a 1$\sigma$ lower limit of 1967 keV. Meanwhile, both \citet{Panagiotou_2020} and this work get $\Gamma \sim$ 1.9. \citet{Pahari_2017} reported an \ec = $143_{-40}^{+96}$ keV and $\Gamma \sim 1.8$. Note \citet{Pahari_2017} employed the quasi-simultaneous Swift-XRT data (6.5 ks XRT exposure, while 21 ks of \Nu) and adopted a quite complex model, which may explain the discrepancy. \citet{Akylas_2021} reported an \ec = $115_{-37}^{+95}$ keV, $\Gamma \sim 1.6$ and $R_{pexmon} \sim 0.74$. The reason behind the discrepancy between \citet{Akylas_2021} and our result (both fitting only NuSTAR spectra) remains unclear and we can not reproduce their result following the same process with the same model of them (the same for NGC 3516 and Ark 120 below).

\item NGC 3516, Sy 1.2, \ec $>$ 476 keV. \citet{Panagiotou_2020} reported a 1$\sigma$ lower limit of 4940 keV.  \citet{Akylas_2021} reported an inconsistent \ec = $89_{-48}^{+24}$ keV and a $R_{pexmon} \sim 1.29$.

\item WKK 1263 (IGR J12415-5750), Sy 1.5, \ec $>$ 530 keV. \citet{Kamraj_2018}, \citet{Panagiotou_2020} and  \citet{Akylas_2021} reported lower limits of 224 keV, 1826 keV and 282 keV respectively, and all three works derive $\Gamma$ $\sim$ 1.8, similar to our results. \citet{Molina_2019} reported an \ec = $123_{-47}^{+54}$ keV, $\Gamma \sim 1.6$ and a similar $R_{pexrav} < 0.23$. The involvement of the quasi-simultaneous Swift-XRT data (5.7 ks XRT exposure, while 16 ks of \Nu) in \citet{Molina_2019} may account for such discrepancy. 

\item Mrk 279, Sy 1.5, \ec $>$ 542 keV. Not reported elsewhere. 

\item MCG-06-30-015, Sy 1.9, \ec $>$ 707 keV. \citet{Panagiotou_2020} reported a 1$\sigma$ lower limit of 12000 keV.

\item Ark 120, Sy 1, \ec $>$ 744 keV. Consistently, \citet{Panagiotou_2020} reported a 1$\sigma$ lower limit of 1631 keV;   \citet{Hinkle_2021} reported an \ec = $506_{-200}^{+814}$ keV;  \citet{Nandi_2021} reported an $T_{\rm e}$ = $222_{-107}^{+105}$ keV;  \citet{Marinucci_2019}  reported an $T_{\rm e}$ = $155_{-55}^{+350}$ keV. The only statistically inconsistent result comes from \citet{Akylas_2021} which reported an \ec = $233_{-67}^{+147}$ keV.

\item NGC 4593, Sy 1, \ec $>$ 800 keV. \citet{Zhangjx2018}, \citet{Panagiotou_2020} and \citet{Akylas_2021} reported an \ec lower limits of 450 keV, 6972 keV and 220 keV respectively. \citet{Ursini_2016} reported an \ec = $470_{-150}^{+430}$ keV. 

\item NGC 4051, Sy 1.5, \ec $>$ 800 keV. \citet{Akylas_2021} reported an \ec lower limits of 846 keV. 
\end{enumerate}

\par In general, our large lower limits to \ec are consistent with most of those from the literature. Discrepancies do exist in some sources, mostly due to the inclusion of the data from other missions in some literature studies. In this work, the $E_{\rm cut}$/$T_{\rm e}$ of the radio-loud and radio-quiet sample are measured with solely \Nu spectra, uniformly processed and analyzed. We therefore anticipate the comparison between two samples in this work is unbiased, though the specific measurement of \ec in individual sources could be altered if including quasi-simultaneous observations or using a more complex model. 
The spectra and the best-fit data-to-model residuals of these ten sources (as shown in the Appendix) have been visually examined and no clear systematical residuals could be identified.

\subsection{The reliability of large \ec}
\par The large \ec lower limits reported in this work, and the generally consistent results from 
literature studies, 
appear to contradict our intuition as such large \ec lower limits are far beyond the \Nu spectral coverage (3--78 keV).
For instance, the correcting factor of an 800 keV exponential cutoff to a single power law is only $\approx e^{-0.1}$ (around 10\%) at 78 keV 
, making the measurements of large \ec only possible in a few brightest sources with sufficiently high \Nu spectral S/N at high energy end.
However, as \citet{Garcia_2015} pointed out, the reflection component, which is sensitive to the spectral shape of the hardest coronal radiation, may assist the measurements of high \ec with \Nu spectra
. Based on the $relxill$ model, they showed that \ec can be constrained at as high as 1 MeV for bright sources.
Below we also demonstrate the effect of the reflection component in model $pexrav$ with spectral simulations.
Using the \Nu spectra of NGC 4051 as input, with \ec set at $10^6$ keV and other parameters at the best-fit values, we generate artificial spectra assuming different $R$ in $pexrav$ using $fakeit$. Fitting the artificial spectra following the same process we apply on the real spectra, we successfully constrain the \ec lower limit to be above 800 keV in 0.2\%, 23\% and 40\% of the mock spectra, for $R =$ 0, 1 and 2, respectively. This clearly shows that large \ec can be better constrained in spectra with stronger reflection component.

\par We also check up other factors which may affect the reliability of the high \ec lower limits. 
The \Nu images have been visually double-checked and confirmed to be normal. Moreover, using the traditional method of background subtraction instead of employing the NUSKYBGD, i.e., extracting the background within a region close to the source, would not alter the main results here.

\begin{figure*}
\centering
\subfloat{\includegraphics[width=0.48\textwidth]{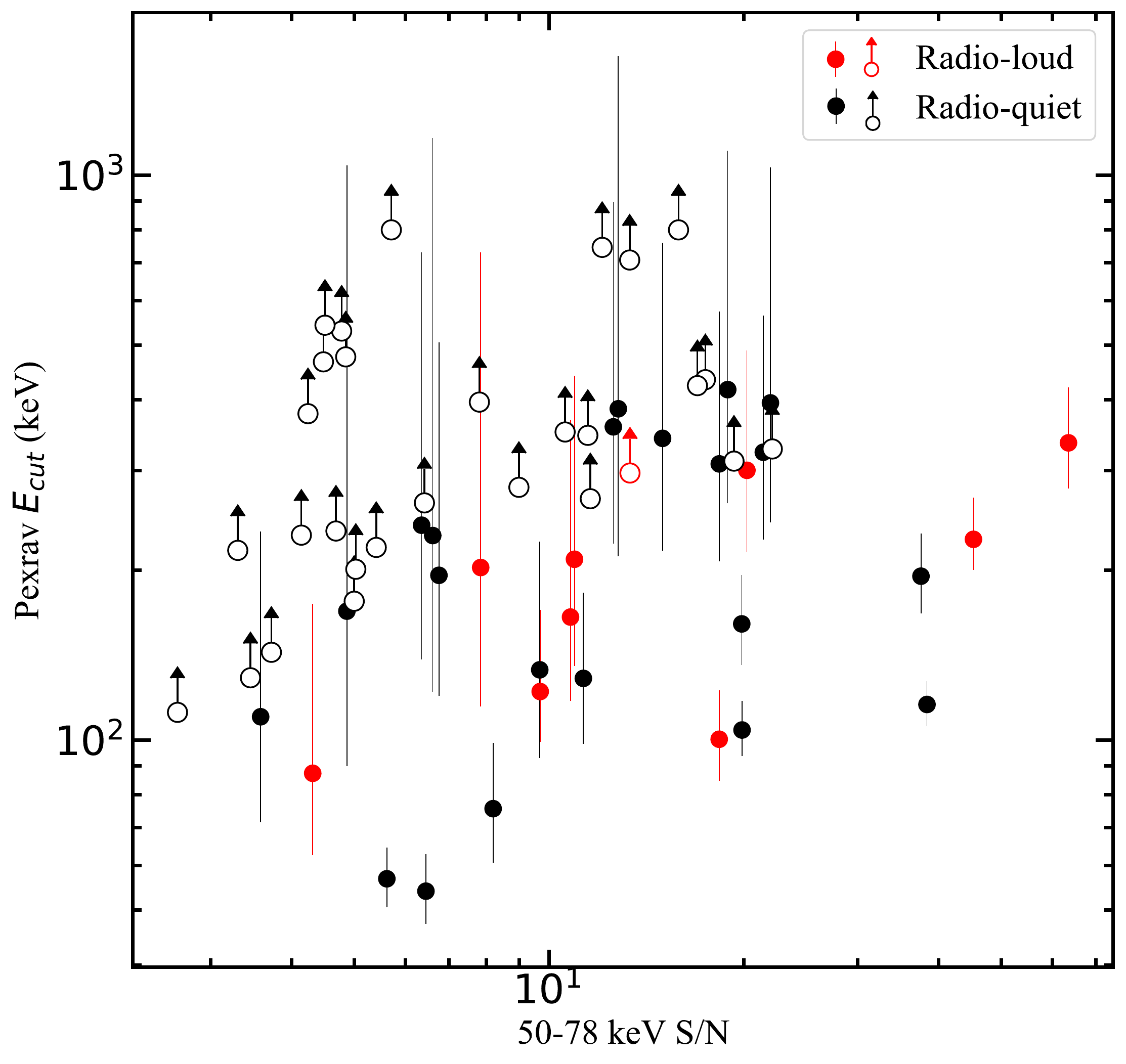}}
\subfloat{\includegraphics[width=0.48\textwidth]{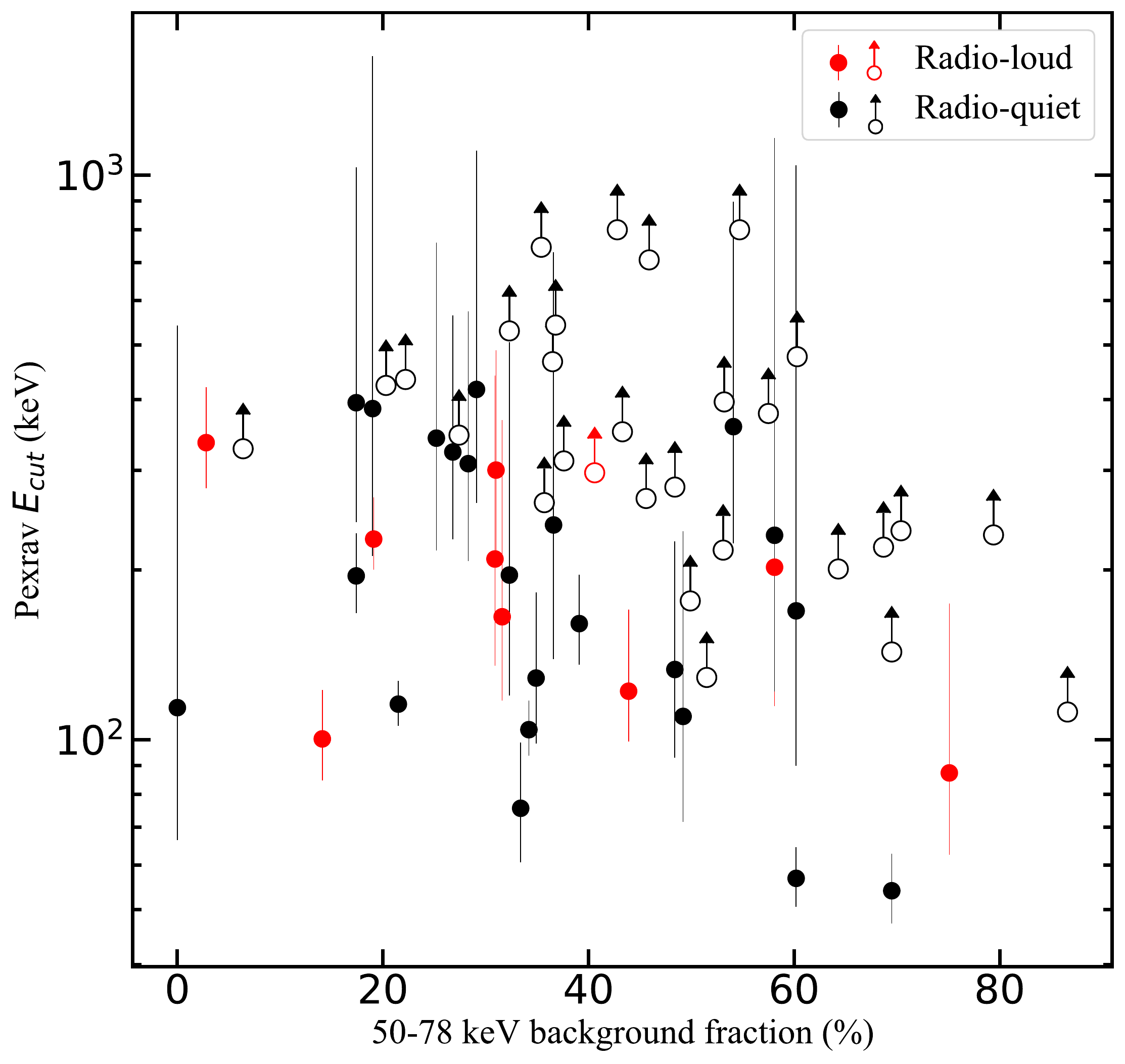}}
\caption{$pexrav$ \ec versus 50--78 keV (FPMA) S/N and background fraction. The S/N would be elevated by a factor of $\sqrt{2}$ if considering both FPMA an FPMB, but the background fraction would remain unchanged. }
\label{fig:backfraction}
\end{figure*}

In Fig. \ref{fig:backfraction} we plot \ec versus the 50--78 keV S/N and \ec versus the 50--78 keV background fraction for our sample. Although in a considerable fraction (32\%) of our sources, their NuSTAR spectra appear background dominated at $>$ 50 keV (i.e., with 50--78 keV background fraction $>$ 50\%), in all but one sources the net 50--78 keV (FPMA) S/N are $>$ 3. This indicates our spectral fitting results are unlikely biased by poor spectral quality or high background level at high energies. From Fig. \ref{fig:backfraction} we also see that $E_{cut}$ lower limits increase with 50--78 keV S/N, and decrease with 50--78 keV background fraction. In other words, these high $E_{cut}$ lower limits ($>$ 400 keV) can only be obtained at relatively higher 50--78 keV S/N and lower 50--78 keV background fraction. This confirms these high $E_{cut}$ lower limits are not due to strong background or poor spectral quality at highest energies.

\begin{figure*}
\centering
\subfloat{\includegraphics[width=0.33\textwidth]{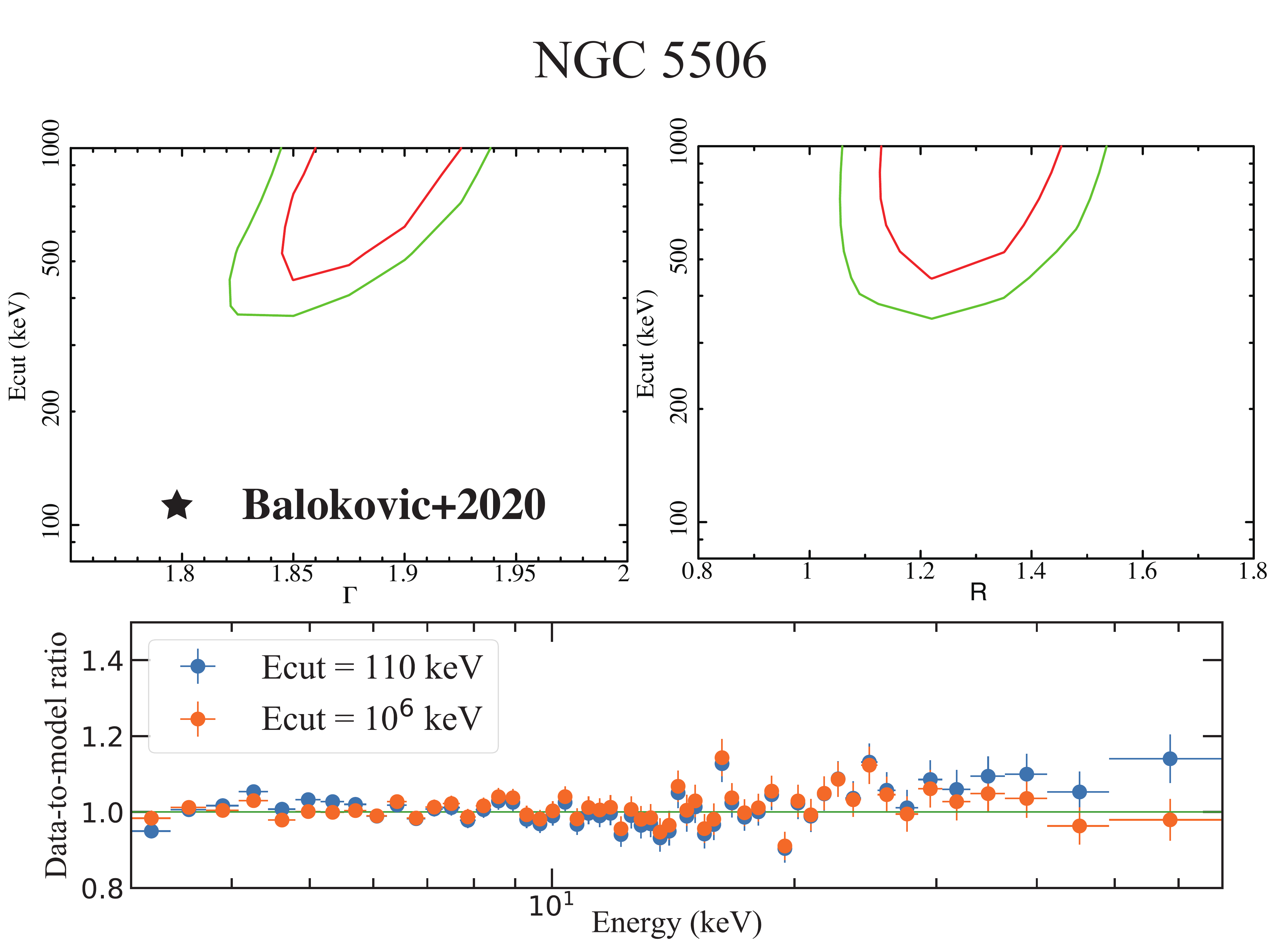}}
\subfloat{\includegraphics[width=0.33\textwidth]{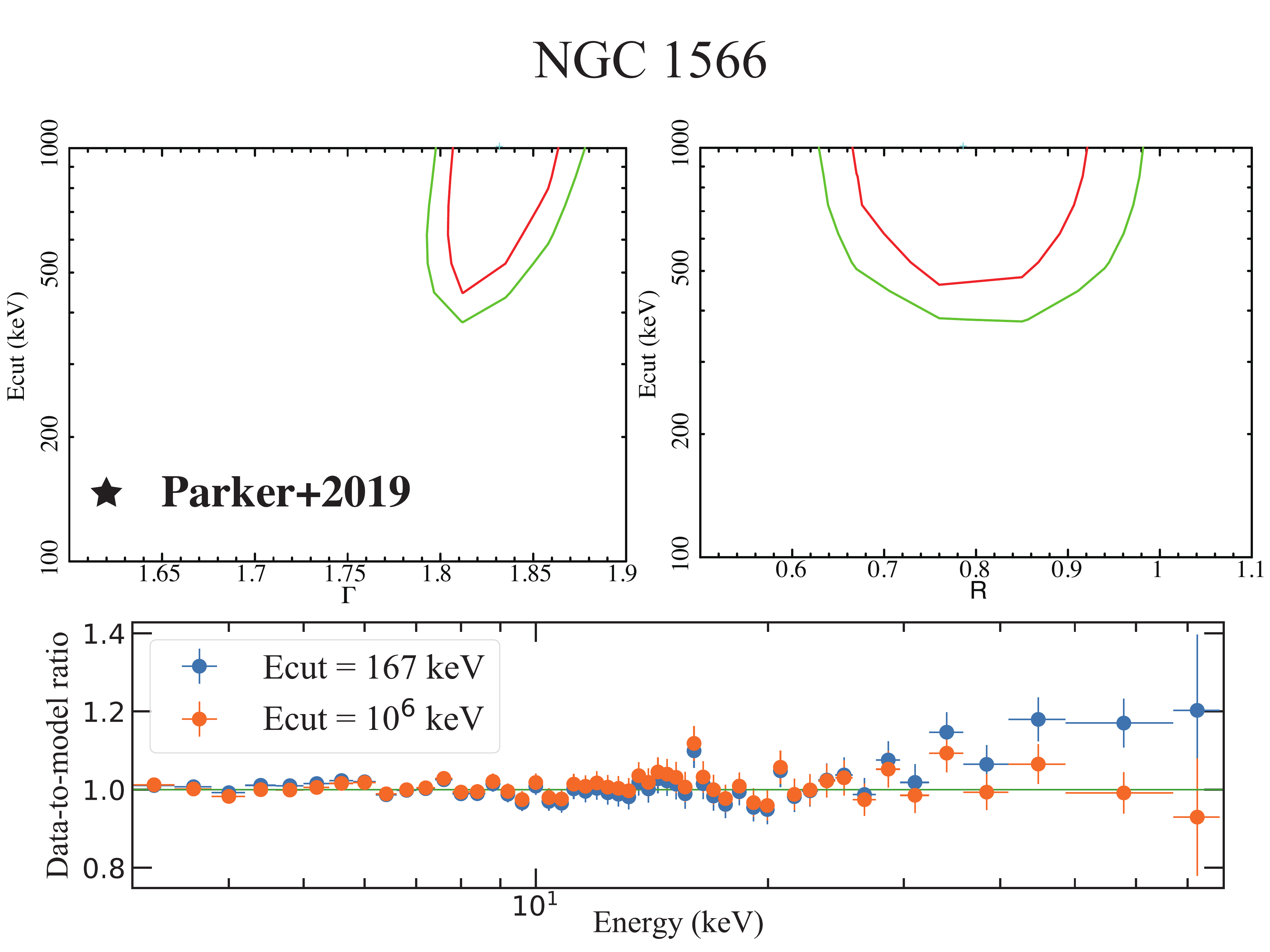}}
\subfloat{\includegraphics[width=0.33\textwidth]{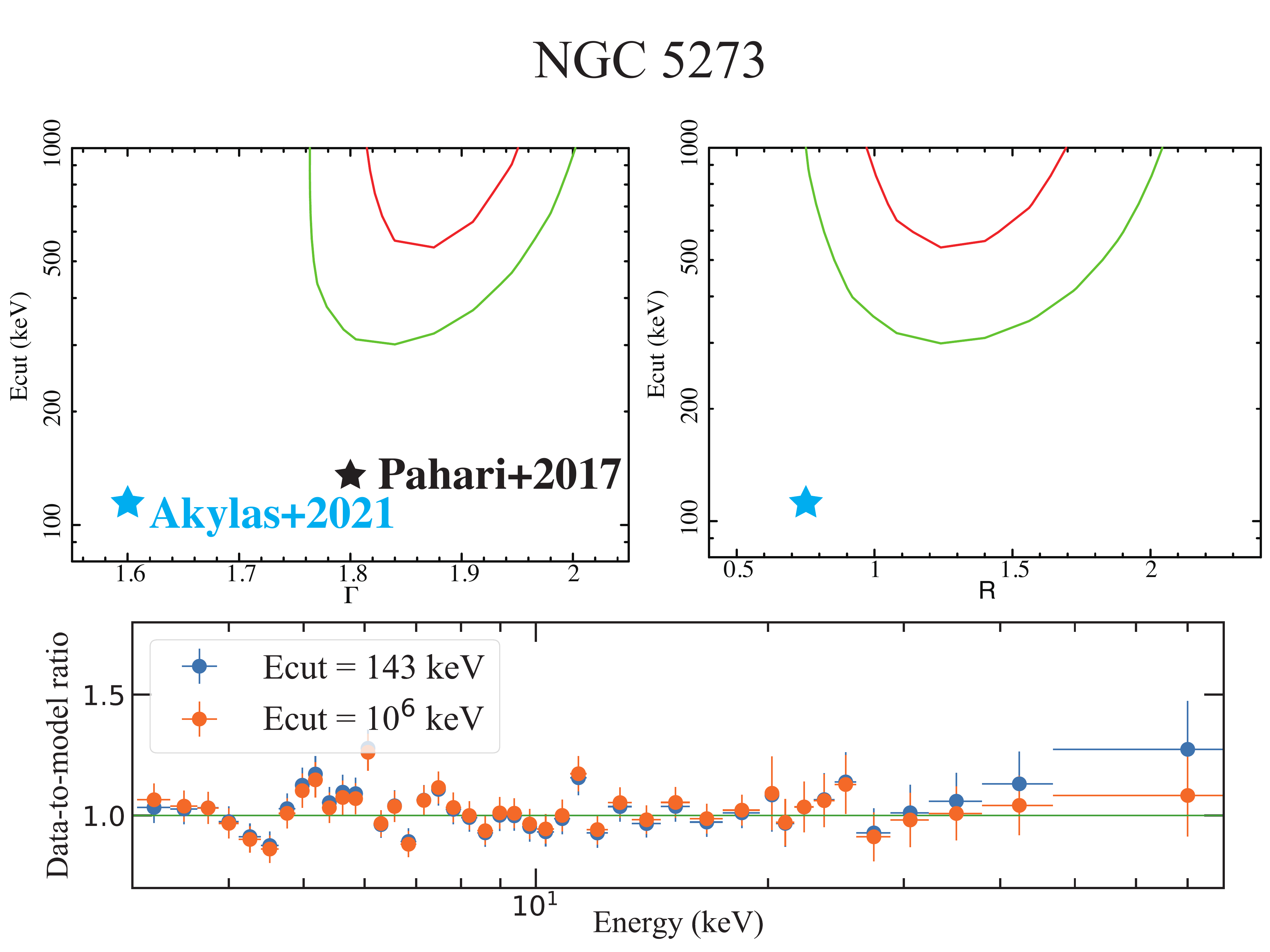}}\\ 
\subfloat{\includegraphics[width=0.33\textwidth]{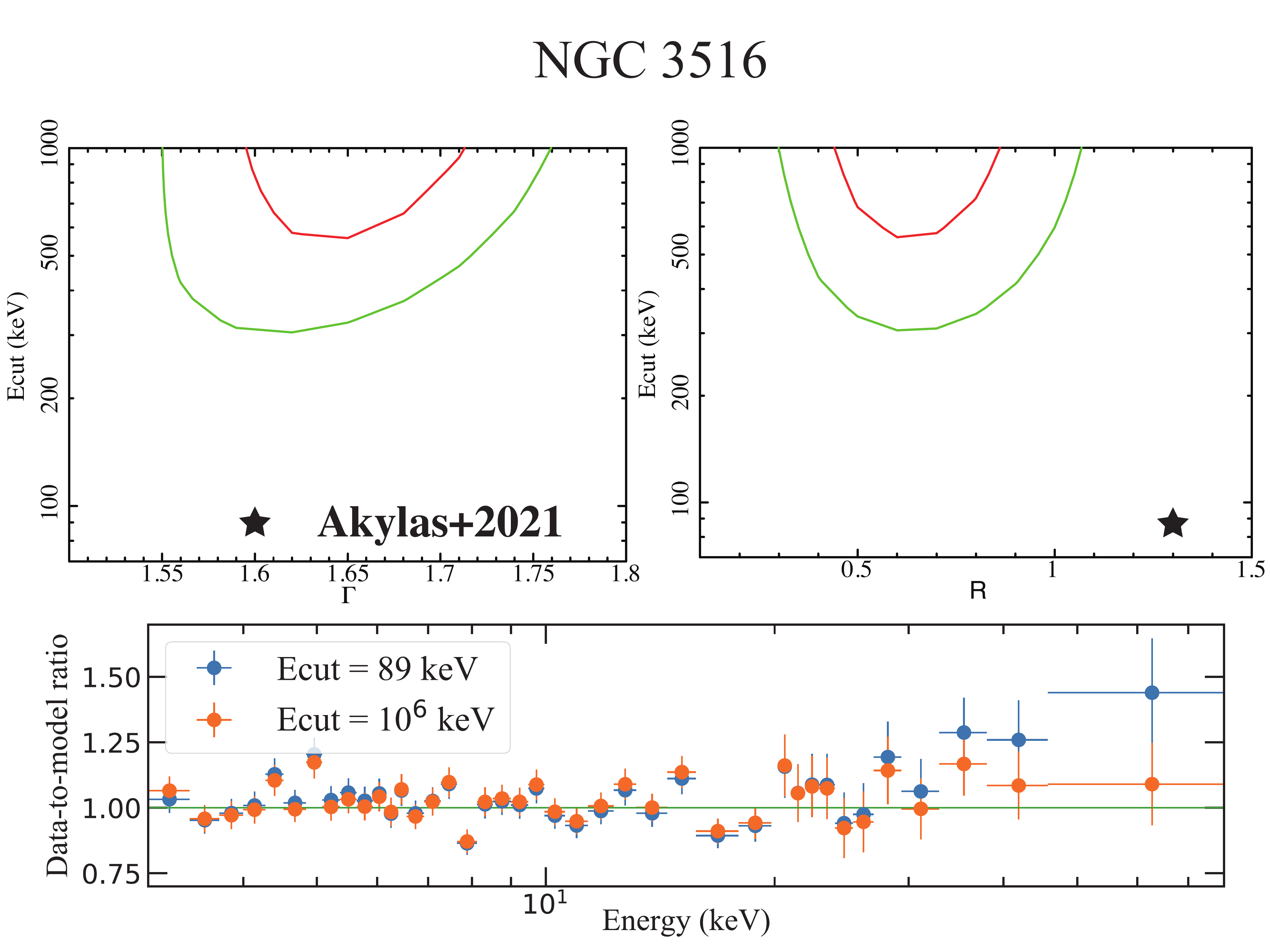}}
\subfloat{\includegraphics[width=0.33\textwidth]{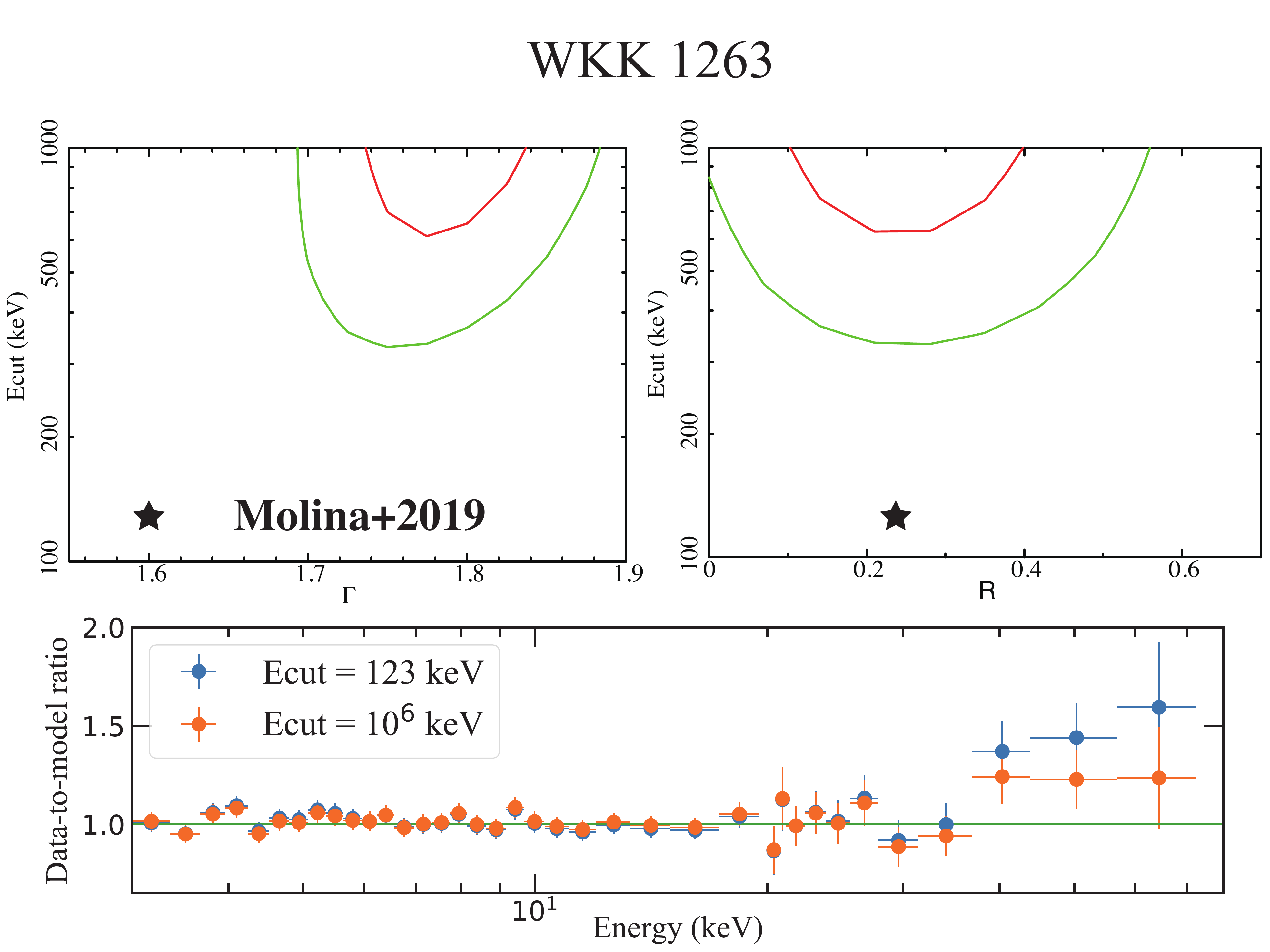}}
\subfloat{\includegraphics[width=0.33\textwidth]{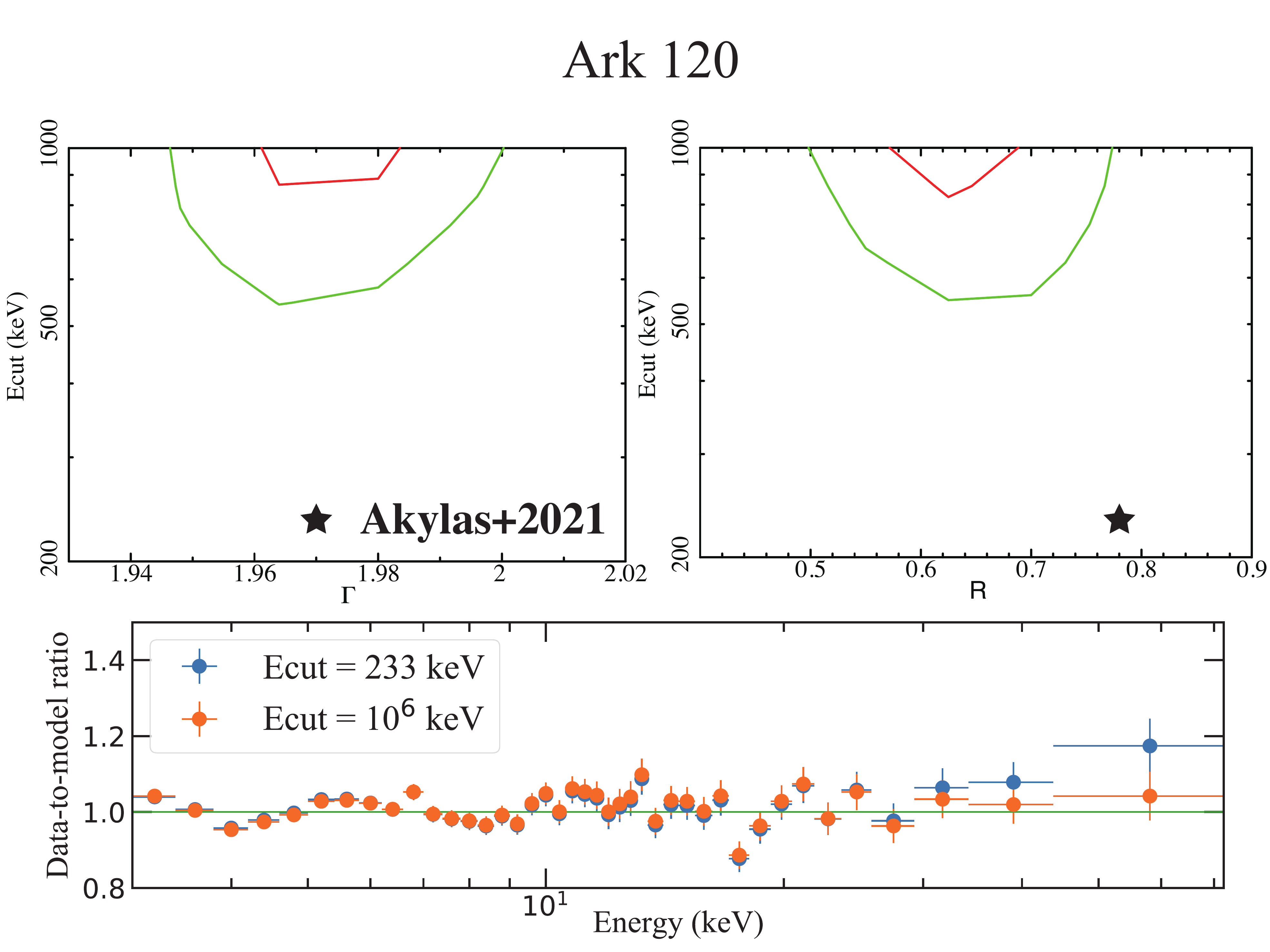}}
\caption{Upper panel: the $\Gamma$ -- \ec and $R$ -- \ec contours (with confidence levels of 1$\sigma$ and 2$\sigma$ plotted, corresponding to $\Delta \chi^2 =$ 2.3 and 4.21) of six sources with \ec lower limits $>$ 400 keV but with statistically inconsistent low \ec detections reported in literature. We mark the reported small \ec detections, together with powerlaw index $\Gamma$ and $pexrav$/$pexmon$ reflection parameter $R$ (when available) from literature for comparison. Lower panel: the data-to-model ratios of the best-fitting results with \ec fixed at $ 10^6$ keV and at the reported low values from literature (see \S\ref{sec:literature}) respectively. For better illustration, the data have been rebinned and only the FPMA spectra are plotted. }
\label{fig:example}
\end{figure*}

Beside, we note 
complex parameter degeneracy may exist between \ec and other parameters \citep[e.g.,][]{Hinkle_2021}.  We hence review the fitting results in individual sources using two parameter contours, among which six sources with \ec lower limits $>$ 400 keV but controversial \ec\ detections\footnote{To highlight the discrepancies between our and literature results for these six sources, in Fig. \ref{fig:example} we also mark the reported statistically-inconsistent \ec detections in literature, together with measurements of powerlaw index $\Gamma$ and reflection parameter $R$ (when available). We clearly see that, even considering two parameter confidence contours, our fitting results statistically challenge those low \ec detections reported in literature. We note those low \ec detections reported in literature are often (in 5 sources) accompanied by spectral indices flatter than our measurements, meanwhile the comparison between our $R$ and literature measurements does not reveal a clear trend.} reported in literature are presented in Fig. \ref{fig:example}. For these six sources, the degeneracies between $E_{\rm cut}$, $\Gamma$ and $R$ are found to be weak, with a 2$\sigma$ \ec lower limit $\sim$ 300 keV obtained even using two parameter confidence contours. 
In addition we also demonstrate how the low \ec detections reported in literature (see \S\ref{sec:literature}) deteriorate the spectral fitting in the lower panel for each source of Fig. \ref{fig:example}. 
We conclude our results are robust in the sense of fitting statistics.
See \S\ref{sec:samplediff} and \S\ref{sec:mechanism} for further discussion on the effect of parameter degeneracy.

\par Finally, the three models we adopted in this work ensure the main results are model independent. As shown in Tab. \ref{tab:fitting_result}, the \ec measurements of $pexrav$ generally agree with those of $relxill$, particularly for those sources with large \ec lower limits. As for the Comptonization model, $T_{\rm e}$ is often harder to be constrained (more lower limits, less detections) than the $E_{\rm cut}$, and the lower limits to $T_{\rm e}$ are smaller than 1/3 \ec \citep{Petrucci_2001} in some sources. This is likely because the e-folded power law produces a smoother break (thus extending to lower energy range and could be better constrained in case of large $T_{\rm e}$/\ec) than Comptonization models \citep{Zdziarski_2003, Fabian_2015}.
However, the overall results from the three models are accordant, i.e., sources with extremely large \ec lower limits do have relatively high $T_{\rm e}$, especially compared with radio-loud sources (see next section). We hence rule out the possibility that the large $E_{\rm cut}$/$T_{\rm e}$ lower limits we obtained are due to unknown faults of certain models.

\subsection{The difference between radio-quiet and loud samples}\label{sec:samplediff}

\par 
We have shown that our radio-quiet sample has considerably larger mean $T_{\rm e}$/\ec compared with the radio-loud one.
To explore the statistical reliability of the difference, we need to explore various biases behind the \ec measurements which might be significant here. The first is the complex degeneracies between the spectral parameters; although we have shown above an example that the degeneracies appear weak in individual sources, we need to quantitatively explore whether such effects could be responsible for the different \ec between two samples. 
The second fact is the radio-loud sample is known to have prominently flatter spectra and weaker reflection component than the radio-quiet one (see Fig. \ref{fig:RYEC}); while flatter spectra imply relatively more photons at high energy end, facilitating the \ec measurement, the weaker reflection could contrarily make it hard to constrain high $E_{\rm cut}$. 
The measurements of \ec also rely on the spectral S/N as shown in Fig. \ref{fig:SNcut}, the effect of which could vary from source to source.
Last but might be most important, the Kaplan-Meier estimator itself can be sensitive to the size of the sample, the fraction of the censored data (lower limits), and the extremely large lower limits. As shown in Fig. \ref{fig:SNcut} and especially in the right panel of Fig. \ref{fig:ETSN}, the mean values, even calculated in the logarithm space, are severely biased towards those large lower limits\footnote{Using median instead of mean hardly improves the situation here, as median is also derived by the estimated probability distribution function when lower limits make up the majority.}. 

\par
\begin{figure}
\centering
\subfloat{\includegraphics[width=0.45\textwidth]{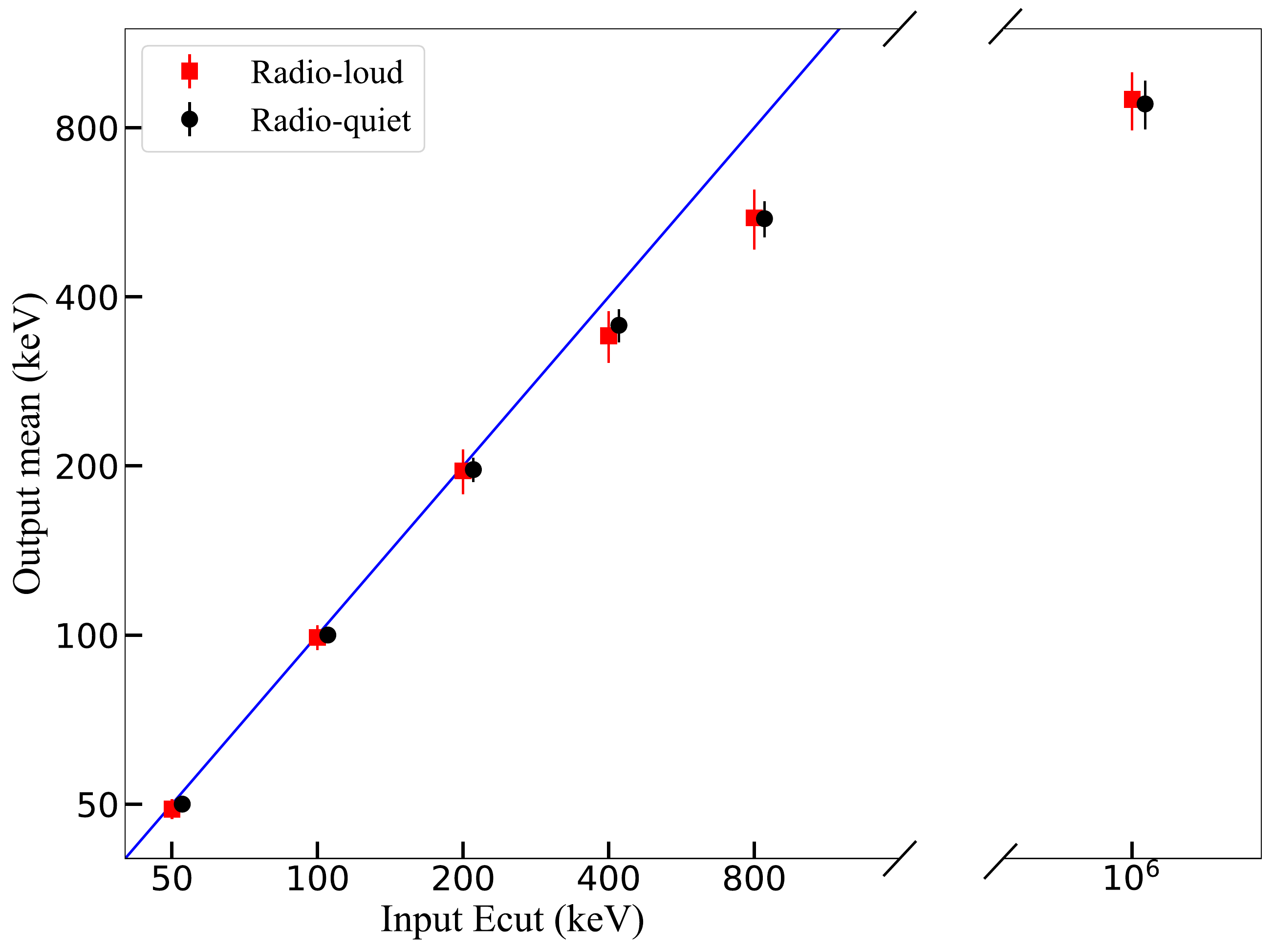}}
\caption{Output mean $E_{\rm cut}$ (in unit of keV) from the mock samples. Note the output mean \ec saturates at $\sim$ 800 keV, partially because we manually set larger \ec lower limits yielded from spectral fitting to 800 keV. \label{fig:fake} }
\end{figure}

\begin{figure}
\centering
\subfloat{\includegraphics[width=0.45\textwidth]{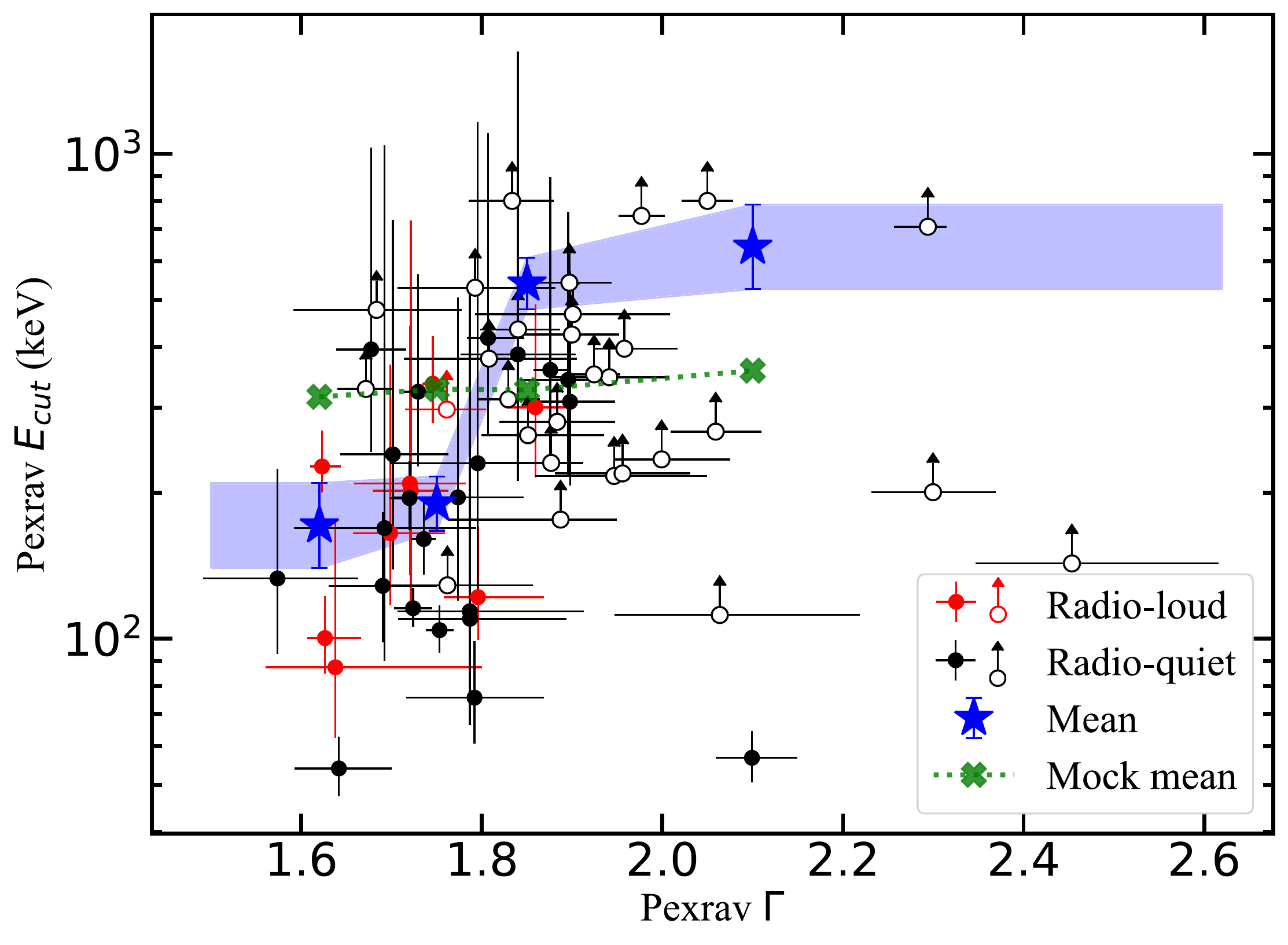}}\\
\subfloat{\includegraphics[width=0.45\textwidth]{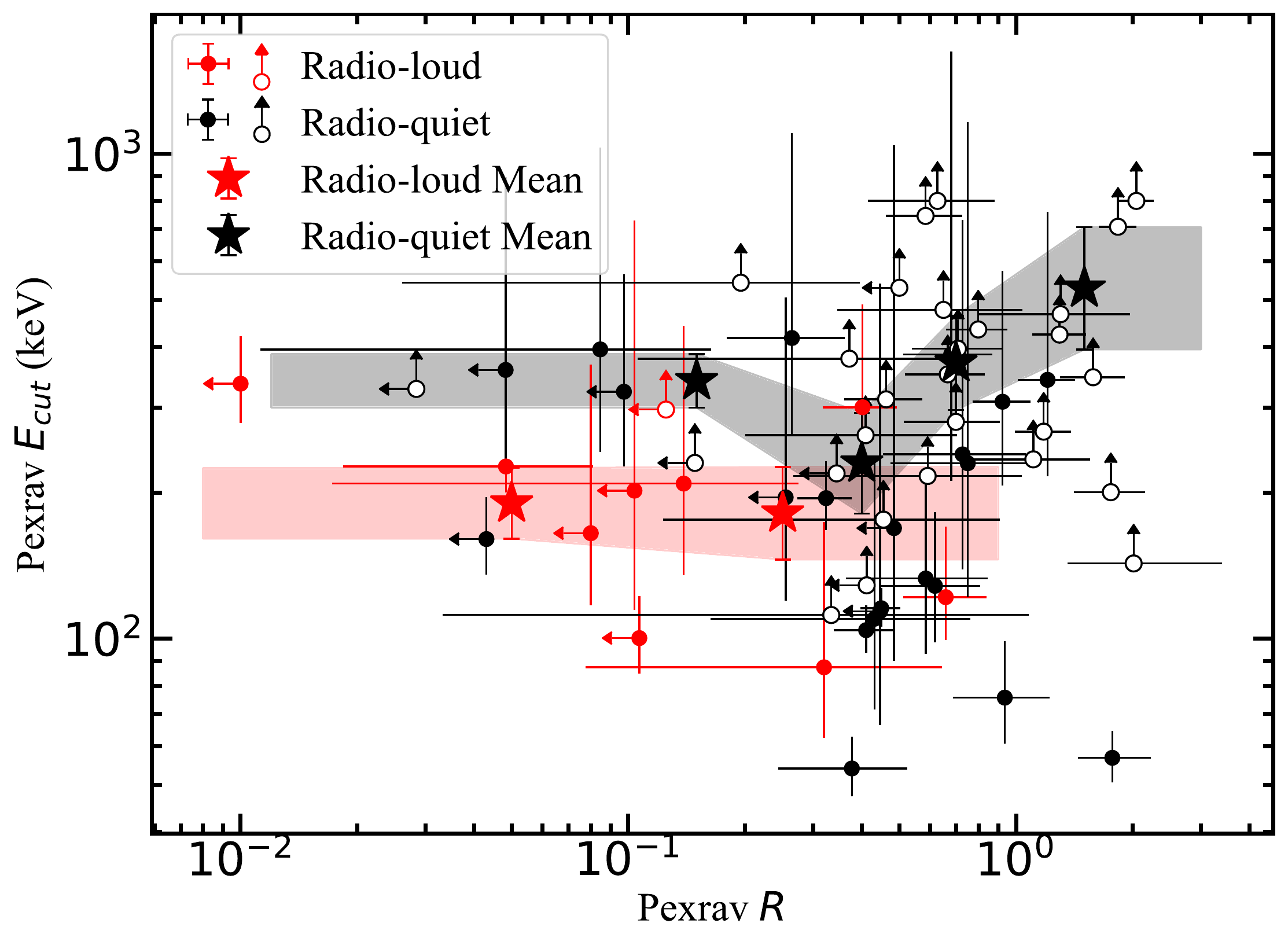}}
\caption{Upper: \ec versus $\Gamma$.  We over-plot the mean \ec within several bins of $\Gamma$ for all sources as we find no difference in the mean \ec between radio-quiet and radio-loud sources at given $\Gamma$. Meanwhile, the output mean \ec versus output $\Gamma$ derived from the mock spectra of the sample (with input \ec = 400 keV) is over-plotted as green crosses.
Lower: \ec versus $R$.     
}
\label{fig:RYEC}
\end{figure}

 To address the overall complicated biases, 
we employ the $fakeit$ within XSPEC to create simulated spectra for each source 
using the best-fit results from $pexrav$\footnote{This whole process is quite computer time consuming, so for simplicity we only perform with the $pexrav$ model.} but manually assigning a set of \ec as input.
We repeat the spectral fitting to the mock spectra and then the measurement of  mean \ec for the mock samples with the Kaplan-Meier estimator, to examine whether our overall procedures could well recover the input \ec or produce artificial different mean \ec between two samples.
As shown in Fig. \ref{fig:fake}, while the simulations do could recover the input \ec in case of low \ec values, high input \ec values (400 keV and above) are clearly underestimated, because the limited bandwidth of NuSTAR, and because we have manually fixed the larger \ec lower limit to 800 keV.
Since a considerably fraction of radio-quiet sources have rather large intrinsic \ec while none of radio-loud sources does, 
this indicates we may have underestimated the mean \ec for our real radio-quiet sample, further strengthening the difference between two samples we have observed.
However, no statistical difference is found between the mock radio-loud and radio-quiet samples. We therefore conclude the biases aforementioned put together are unable to account for the difference in the \ec distribution between two samples.

\par The larger average \ec in radio-quiet sources is however surprising, 
as we would anticipate larger observed \ec in the radio-loud sample due to potential jet contamination \citep{Madsen_2015_3C273} or the stronger Doppler boosting of an outflowing corona in radio AGNs \citep[e.g.][]{Beloborodov_1999, Liu2014, Kang_2020}, even if two populations have the same intrinsic coronal temperature. 
The key underlying reason might be the different $\Gamma$ distribution of the two samples. In Fig. \ref{fig:RYEC} we plot \ec versus $\Gamma$ and the reflection strength $R$ from $pexrav$ for the two samples.  We find that \ec is positively correlated with $\Gamma$ and those large \ec lower limits are mainly detected in sources with steep spectra. Besides, we find no difference in \ec between two populations at comparable $\Gamma$. Therefore, the difference in \ec between two populations could dominantly be attributed to the fact that \ec correlates with photon index $\Gamma$ while the radio-loud sample is dominated by sources with flat spectra. 
Meanwhile, \ec exhibits no clear correlation with $R$, while RQ AGNs do show larger \ec compared with RL ones at given $R$, which could be attributed to the effect of $\Gamma$. 
We note that 
\citet{Kang_2020} found the \ec distribution of their radio-loud sample 
is indistinguishable from that of a radio-quiet sample from \citet{Rani_2019}.
This is likely because the sample of \citet{Rani_2019} is incomplete, which only collected from literature sources with well-constrained \ec and most lower limits were excluded.
In fact, if we drop lower limits from our samples in this work, we would find no difference either in mean \ec between two populations.
Meanwhile, \citet{Gilli_2007} has shown an average \ec of above 300 keV can saturate the X-ray Background at 100 keV. The fact that large \ec mainly exist in steeper spectra also renders our large mean value of \ec in radio-quiet AGNs compatible with \citet{Gilli_2007}, as sources with steep X-ray spectra make little contribution to the high energy X-ray background even with a large $E_{\rm cut}$. 

\subsection{The underlying mechanisms}\label{sec:mechanism}
\begin{figure}
\centering
\subfloat{\includegraphics[width=0.45\textwidth]{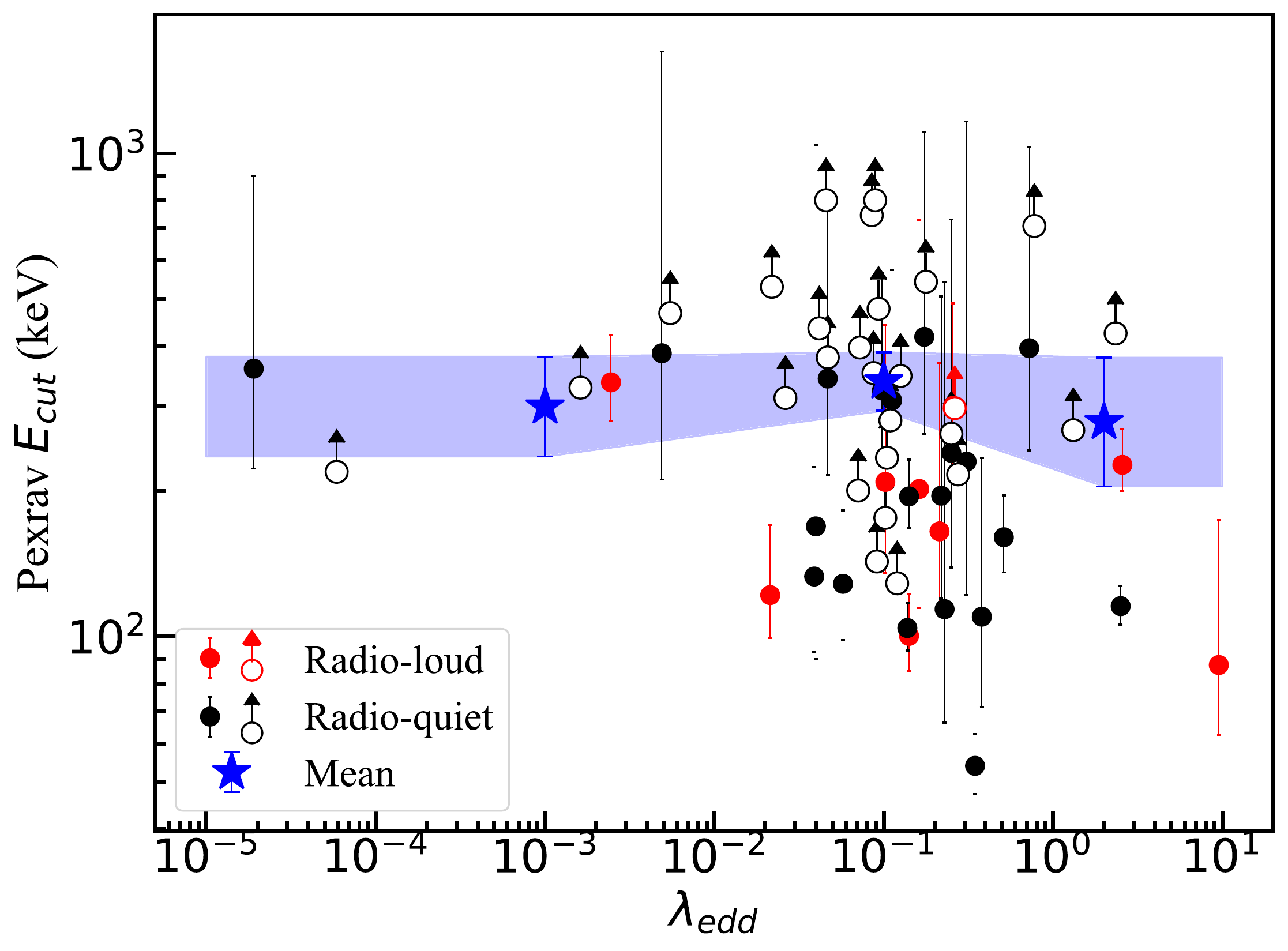}}\\
\subfloat{\includegraphics[width=0.45\textwidth]{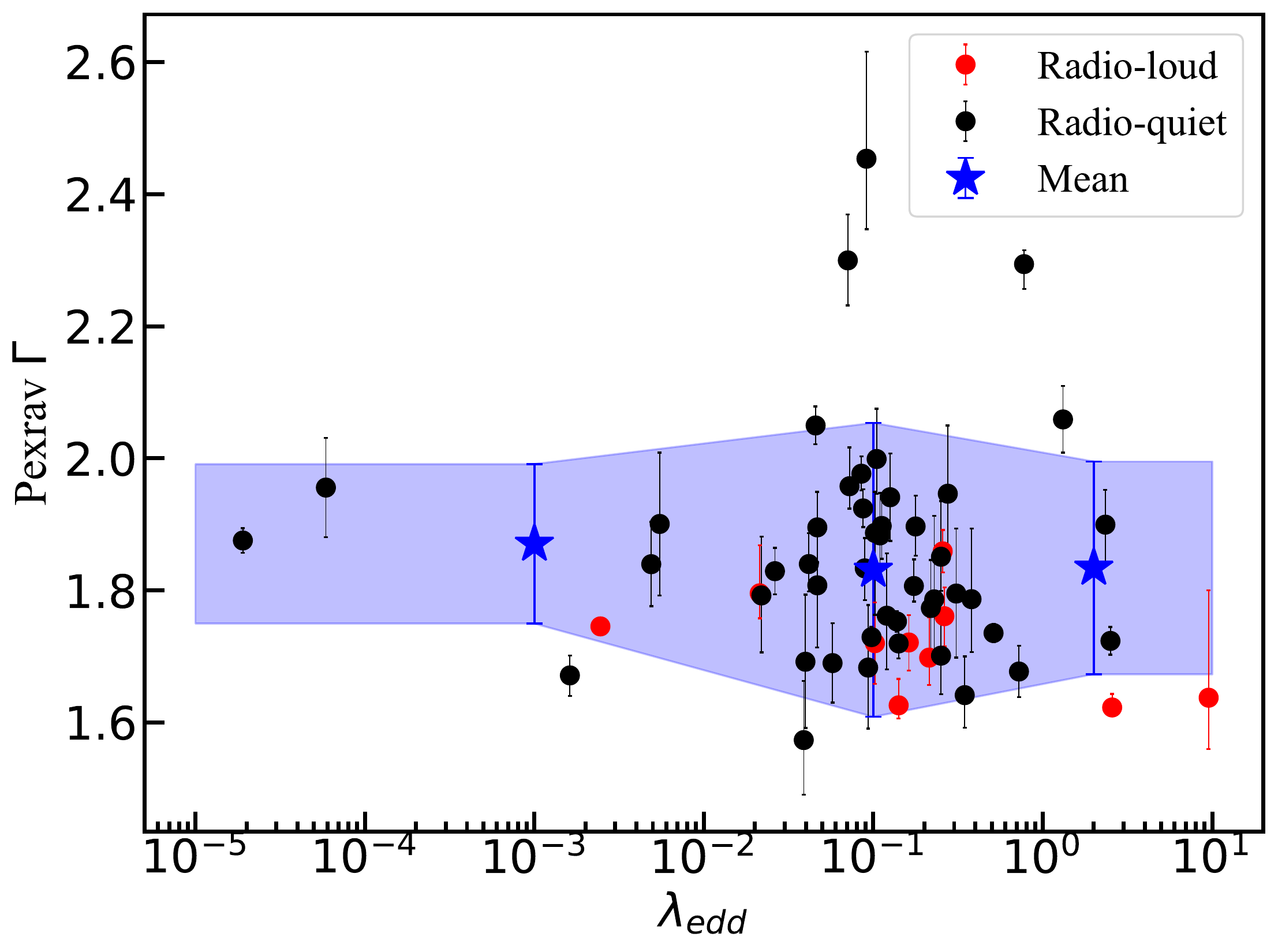}}
\caption{\ec -- $\lambda_{\rm edd}$ and $\Gamma$ -- $\lambda_{\rm edd}$ for our samples. $\lambda_{edd}$ is derived using the black hole mass from literature and up-scaled BAT 14--195 keV luminosity (see Tab. \ref{tab:source_details}).
}
\label{fig:edd}
\end{figure}  

\begin{figure*}
\centering
\subfloat{\includegraphics[width=0.33\textwidth]{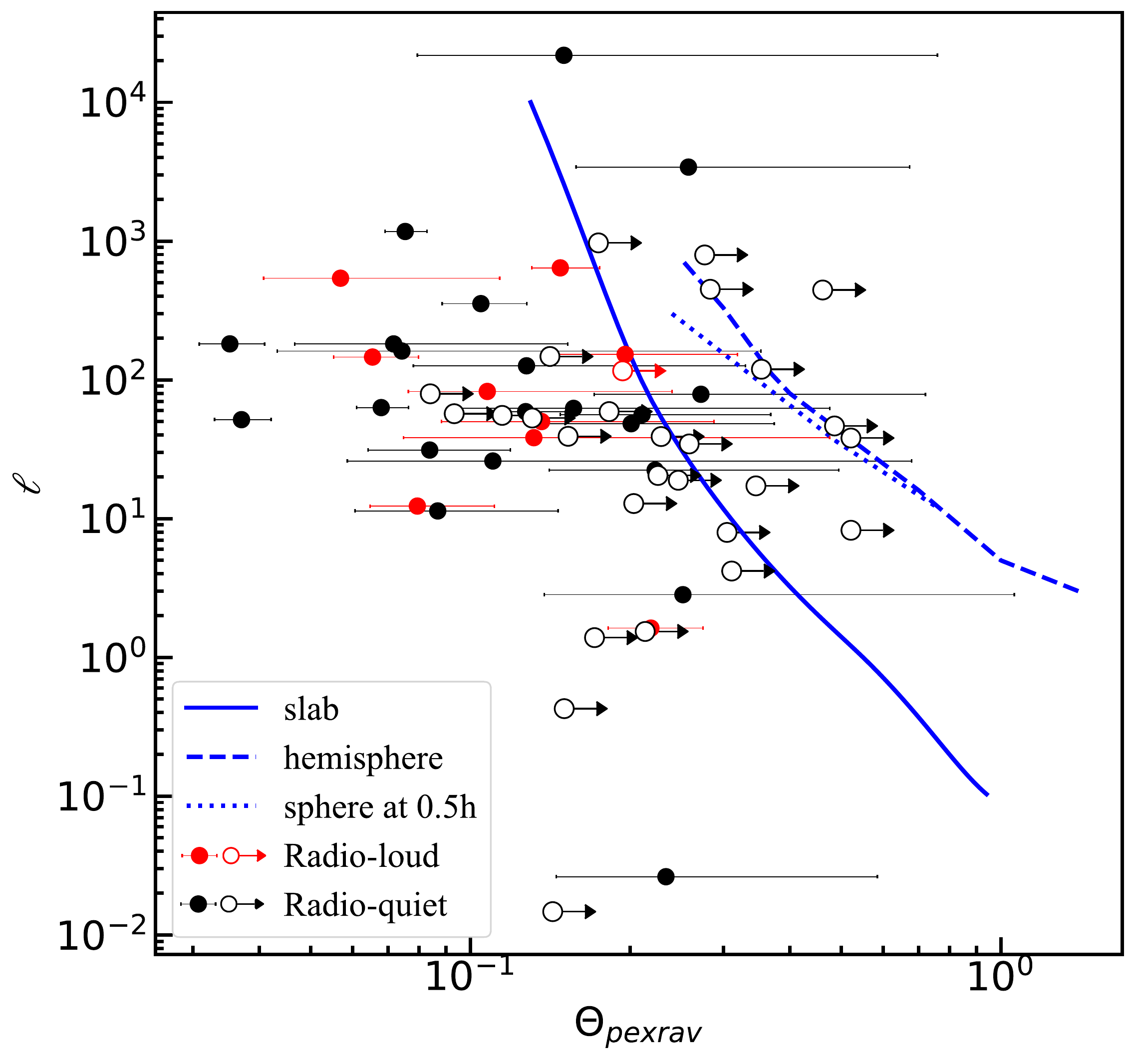}}
\subfloat{\includegraphics[width=0.33\textwidth]{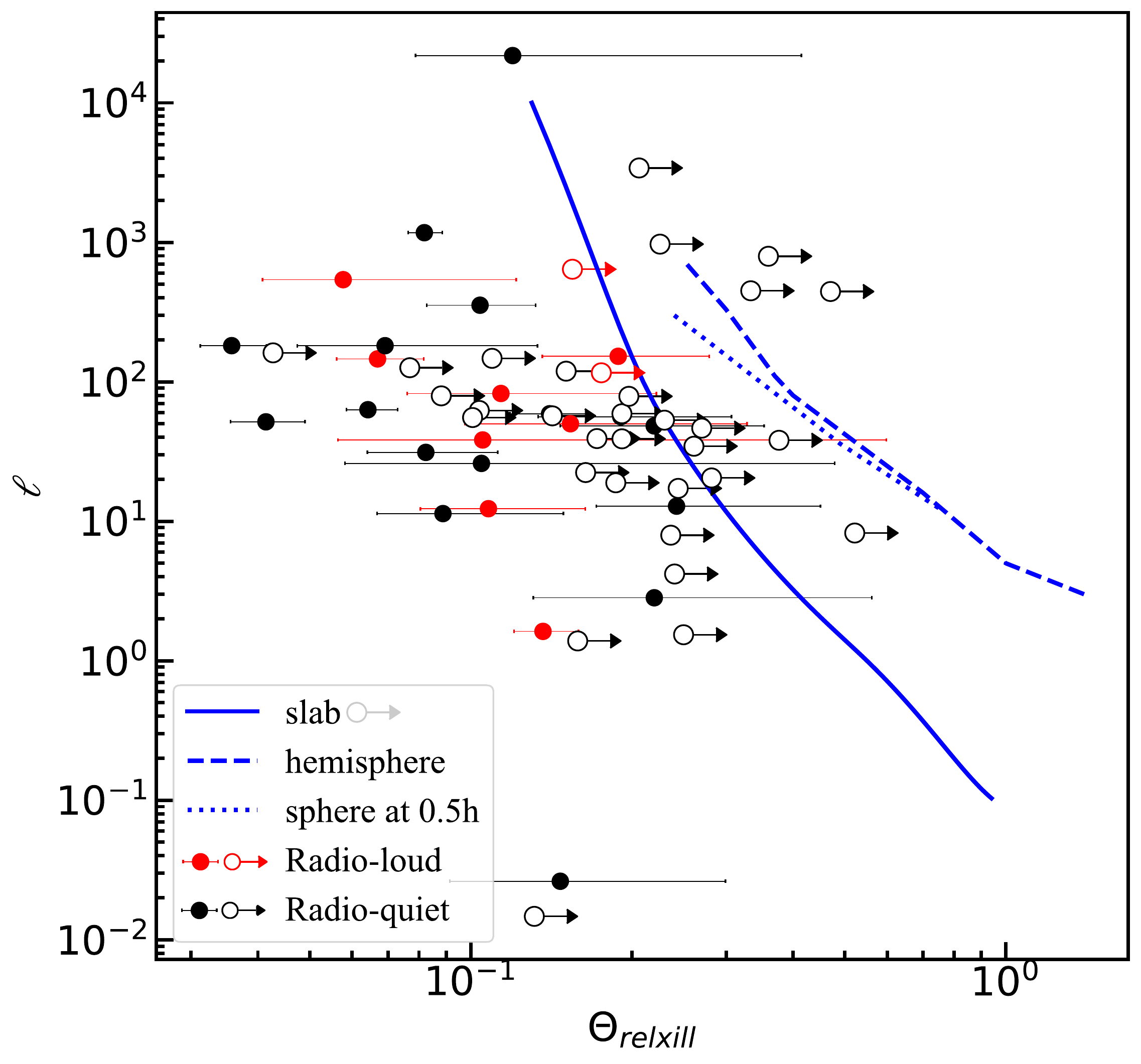}}
\subfloat{\includegraphics[width=0.33\textwidth]{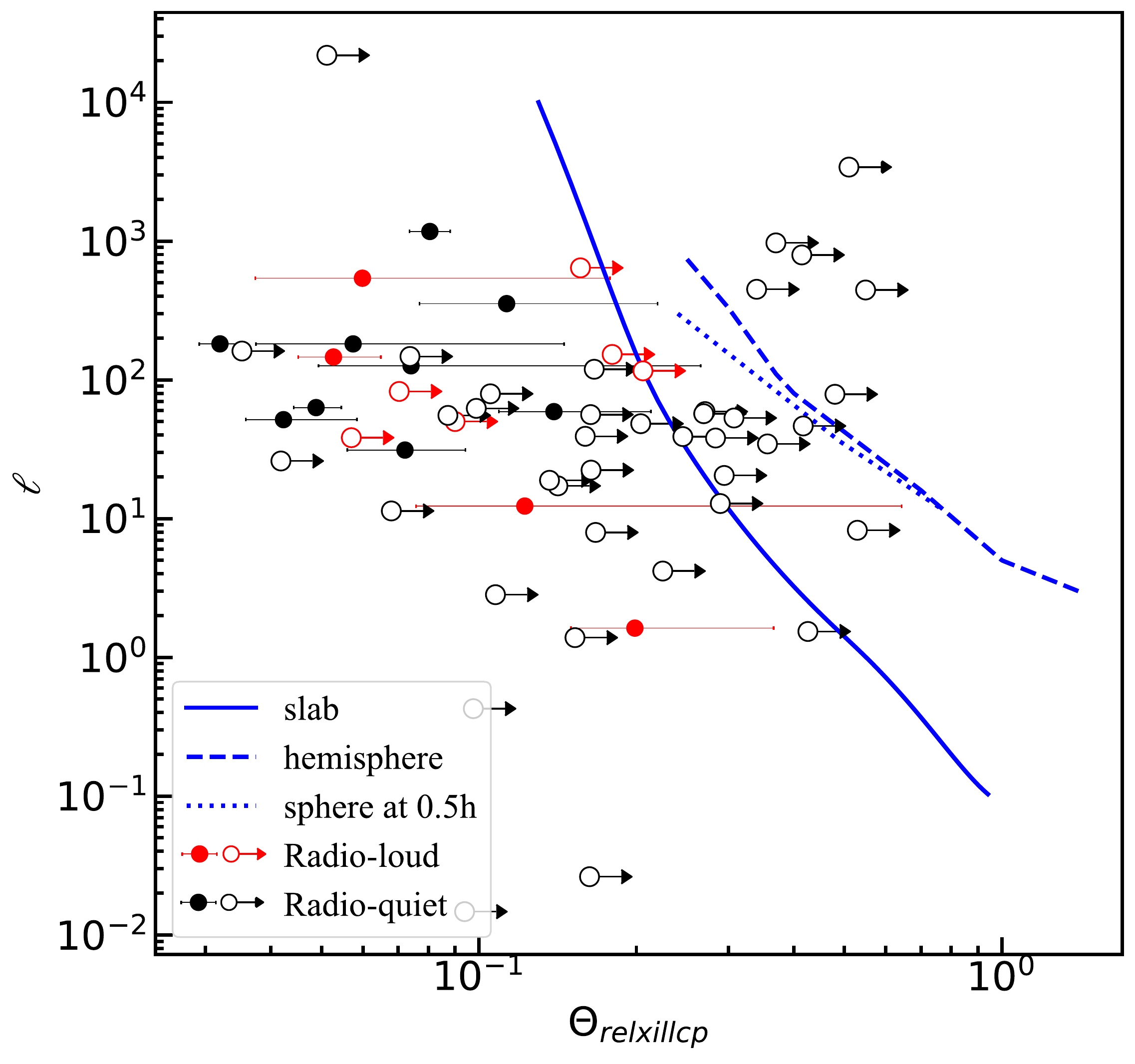}}
\caption{The compactness--temperature ($l$--$\Theta$) diagrams, with $\Theta$ derived from three spectral models. }
\label{fig:compactness}
\end{figure*}

\par Tentative positive correlation between \ec and $\Gamma$ has been reported in other studies with \Nu \citep[e.g.][]{Kamraj_2018, Molina_2019,Hinkle_2021}, and previously with BeppoSAX data \citep[e.g.][]{Petrucci_2001}, however not as pronounced as we have found, likely because of smaller sample size or the domination by poorly constrained lower limits. For instance, the sample in \citet{Kamraj_2018} consists of 46 sources, whereas \ec can be well constrained in only two of them.  The samples in \citet{Molina_2019} and \citet{Hinkle_2021} consist of 18 and 33 sources respectively, considerably smaller than the one presented in this work;
meanwhile, the inclusion of XRT and XMM-Newton data in those two works may have disturbed the measurements of \ec and $\Gamma$ as already discussed above. 

\par The tentative \ec -- $\Gamma$ correlation reported in literature had often been attributed to the parameter degeneracy between \ec and $\Gamma$. 
In this work, the correlation between \ec and $\Gamma$ is rather strong, and $\Gamma$ is well constrained thanks to the high-quality \Nu spectra. We thus expect the effect of such degeneracy to be insignificant. We perform simulations to quantify such effect in our sample. 
Utilizing an \ec=400 keV as input and other best-fit spectral parameters from $pexrav$, we simulate mock spectra for each source. We then examine the correlation between the output \ec and output $\Gamma$ for the mock sample. 
As shown Fig. \ref{fig:RYEC}, while the parameter degeneracy does yield a weak artificial correlation between the output \ec and $\Gamma$, it is much weaker and negligible compared with the observed one.

\par 
The positive correlation between \ec and $\Gamma$ found in this work indicates sources with steeper X-ray spectra tend to hold hotter coronae.
Subsequently, to produce the steeper spectra, the hotter coronae need to have lower opacity. The negative link between coronal temperature and opacity could partly be attributed to the fact that the cooling is more efficient in coronae with higher opacity, i.e., sustainable hotter coronae are only possible with lower opacity. However, while lower opacity could lead to steeper spectra, higher temperature alters the spectral slope towards an opposite direction. While it is yet unclear what drives the positive \ec -- $\Gamma$ correlation reported in this work, it is intriguing to compare it with how \ec varies with $\Gamma$ in individual AGNs. \ec variabilities detected in several individual AGNs show a common trend that when an individual source brightens in X-ray flux, its power law spectrum gets softer and \ec increases, also revealing a positive \ec -- $\Gamma$ correlation \citep[hotter-when-softer/brighter, e.g.][]{Zhangjx2018, Kang_2021}. 
However, the similarity between the two types of positive \ec -- $\Gamma$ correlation (intrinsic: in individual AGNs,  versus global: in a large sample of AGNs) does not necessarily imply common underlying mechanisms. This is because, 
while the intrinsic \ec -- $\Gamma$ correlation, which could be accompanied with dynamical/geometrical changes of the coronae such as inflation/contraction \citep{Wu_2020}, reflects variations in the inner most region of individual AGNs, the global \ec -- $\Gamma$ correlation we find in a sample of AGNs shall mainly reflect the differences in their physical properties, including SMBH mass, accretion rate and other unknown parameters. 
\citet{Kang_2021} also found a tentative trend that \ec reversely decreases with $\Gamma$ at $\Gamma$ $>$ 2.05 in one individual source, yielding a $\Lambda$ shape in the \ec -- $\Gamma$ diagram. Such trend however is not seen in the global \ec -- $\Gamma$ relation. 

\par The positive global \ec -- $\Gamma$ correlation also implies a potential positive correlation between \ec and Eddington ratio $\lambda_{edd}$, as sources with higher accretion rate tend to have steeper spectra \citep[e.g.][]{Shemmer_2006, Risaliti_2009, Yang_2015}.  However, we find no significant correlation between $\lambda_{edd}$ and $\Gamma$, or between $\lambda_{edd}$ and $E_{\rm cut}$ in our sample (see Fig. \ref{fig:edd}), consistent with the results of \citet{Molina_2019}, \citet{Hinkle_2021} and \citet{Kamraj_2022}.
This is likely because the uncertainties in the measurements of $\lambda_{edd}$ are large, or the \ec -- $\Gamma$ correlation we find is not driven by Eddington ratio. 
However, our results disagree with \citet{Ricci_2018}, which claimed a negative correlation between \ec and $\lambda_{edd}$ based on SWIFT BAT spectra. 
But note a dominant fraction (144 out of 212) of the \ec measurements reported in \citet{Ricci_2018} are lower limits\footnote{Besides, we are unable to reproduce the negative correlation given in Fig. 4 of \citet{Ricci_2018} utilizing their data and the approach adopted in this work to estimate the median $E_{\rm cut}$. Instead, we find no clear correlation either between \ec and $\lambda_{edd}$ using their sample and data.}.

\par We note a couple of individual local sources with high Eddington ratios ( $>$ 1) have been reported with \Nu spectra to have low $E_{\rm cut}$/$T_{\rm e}$ in literature  \citep[e.g., Ark 564, IRAS 04416+1215, ][]{Kara_2017, Tortosa2022}, seeming to suggest lower coronal temperature at higher Eddington ratio.
While the \Nu spectral quality of Ark 564 is rather high (10--78 keV FPMA S/N = 88), it is not in the 105-month SWIFT/BAT catalog, thus not included in this work.
The \Nu spectral quality of IRAS 04416+1215 (with 10--78 keV S/N of 10) is much poorer compared with the sample presented in this work, and  
our independent fitting to its \Nu spectra alone could only yield poorly constrained lower limits to its $E_{\rm cut}$ or $T_{\rm e}$.
Utilizing XMM-Newton and \Nu data, low $E_{\rm cut}$/$T_{\rm e}$ is also detected in a high-redshift source with Eddington ratio $>$ 1 \citep[PG 1247+267, ][]{Lanzuisi_2016}. 
However, its NuSTAR spectra also have poor S/N  ($\sim$ 20 in the rest frame 10--78 keV). 
Meanwhile, simply collecting positive detections of $E_{\rm cut}$/$T_{\rm e}$ from literature could suffer significant publication bias. 

\par We finally plot the samples on the well-known compactness--temperature ($l$--$\Theta$) diagram.
\citet{Fabian_2015} has shown that, the AGN coronae locate near the boundary of the forbidden region in the $l$--$\Theta$ diagram, 
suggesting the coronal temperature is governed and limited by runaway pair production.
Following \citet{Fabian_2015}, we calculate the compactness, $l = 4\pi (m_p / m_e)(r_g / r)(L / L_{\rm edd})$, and dimensionless temperature, $\Theta = kT_e / m_e c^2$. We assume a $r = 10$  $r_g$, adopt the unabsorbed 0.1--200 keV primary continuum luminosity extrapolated by the best-fit $pexrav$ model to NuSTAR spectra (listed in Table \ref{tab:source_details}), and calculate $L_{\rm edd}$ using the SMBH mass in Table \ref{tab:source_details}\footnote{Note the $\lambda_{edd}$ presented in Table \ref{tab:source_details} was derived using up-scaled BAT 14--195 keV luminosity, thus the ratio of the compactness parameter (calculated using 0.1--200 keV measured with NuSTAR spectra) to $\lambda_{edd}$ could deviate from a single constant.}.
For $pexrav$ and $relxill$, the $T_{\rm e}$ is approximated by $E_{\rm cut}$/3 \citep{Petrucci_2001}, while for $relxillcp$ the measured $T_{\rm e}$ is directly used. The $l$--$\Theta$ diagrams of the three models are shown in Fig. \ref{fig:compactness}, with the boundaries of runaway pair production of the three geometries \citep{Stern_1995,Svensson_1996} over-plotted. Apparently, the sources in this work have a wider $\Theta$ range compared with that of \citet{Fabian_2015}, likely because of the large sample size of this work. 
On the one hand, there are many sources lying clearly to the left of the slab pair line, particularly sources in the upper left corner in the $l$--$\Theta$ diagram. They appear to support the existence of hybrid plasma in the coronae as hybrid plasma would shift the pair line to the left and the shift is more prominent in the top of the line \citep[see Fig. 6 in][]{Fabian2017}.
On the other hand, both the directly measured and conservatively estimated $T_{\rm e}$ (1/3 $E_{\rm cut}$ here, while 1/2 in \citealt{Fabian_2015}) of a considerable fraction of sources lie beyond (to the right of) the slab pair line, consistent with \citet{Kamraj_2022}, 
favoring the sphere or hemisphere geometry. 
Considering the \ec -- $\Gamma$ relation shown above, it is implied that the coronal geometry might be spectral slope dependent,
i.e., flatter shape for harder spectra, and rounder for softer spectra. 
Furthermore there are several sources with lower limits of $\Theta$ lying even beyond the boundaries of all three geometries, which suggests their coronae could be more extended than 10 $r_g$ we have assumed.

 \acknowledgments
 {This research has made use of the NuSTAR Data Analysis Software (NuSTARDAS) jointly developed by the ASI Science Data Center (ASDC, Italy) and the California Institute of Technology (USA). The work is supported by National Natural Science Foundation of China (grants No. 11890693, 12033006 $\&$ 12192221). The authors gratefully acknowledge the support of Cyrus Chun Ying Tang Foundations. 
 }
 
 \software{HEAsoft (v6.28; HEASARC 2014),  NuSTARDAS, NUSKYBGD \citep{Wik_2014}, XSPEC \citep{Arnaud_1996}, ASURV \citep{Feigelson_1985}, TOPCAT \citep{TOPCAT}, GNU Parallel Tool \citep{Tange2011a}.}
\clearpage

\appendix
\begin{figure*}[h!]
\centering
\subfloat{\includegraphics[width=0.33\textwidth]{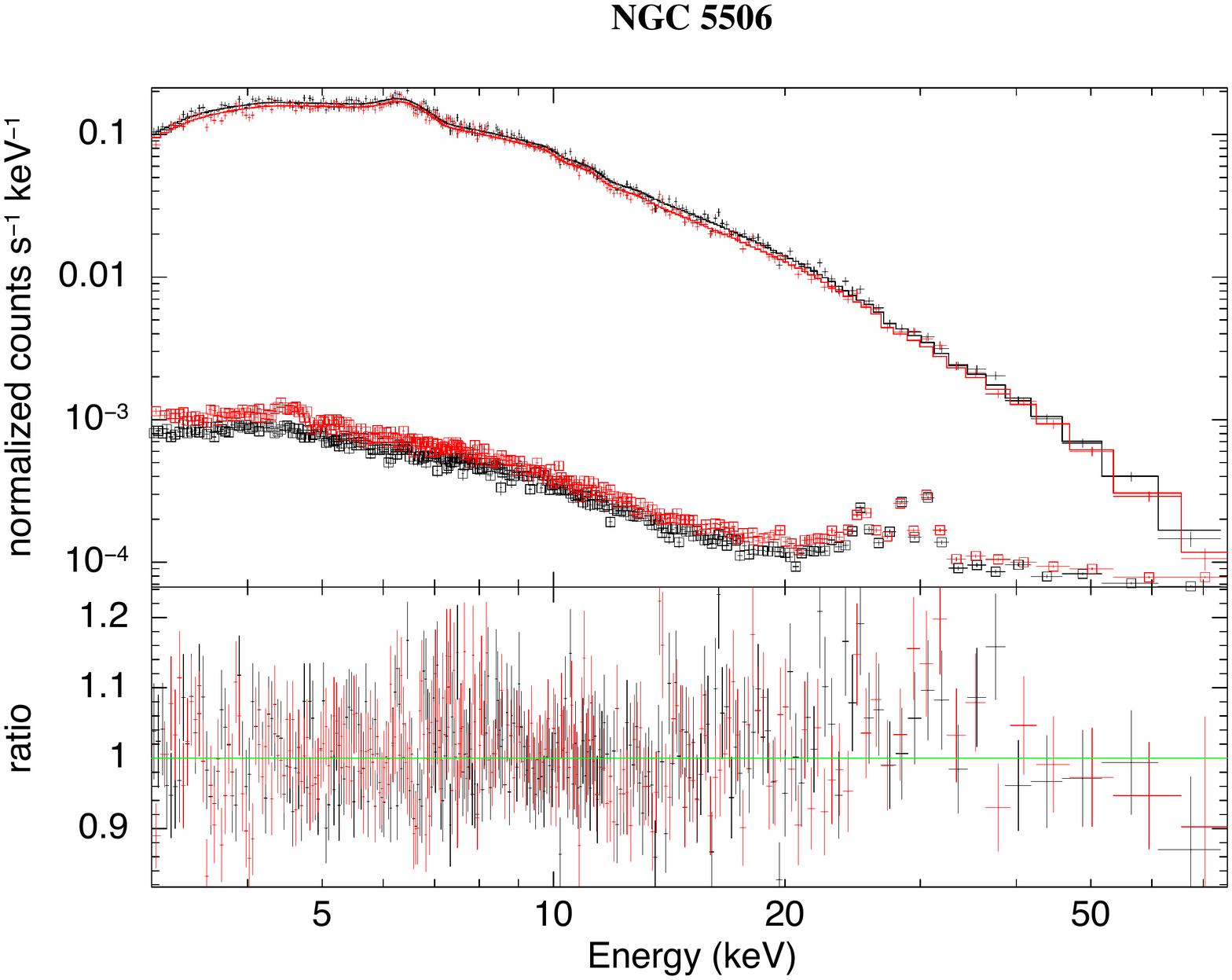}}
\subfloat{\includegraphics[width=0.33\textwidth]{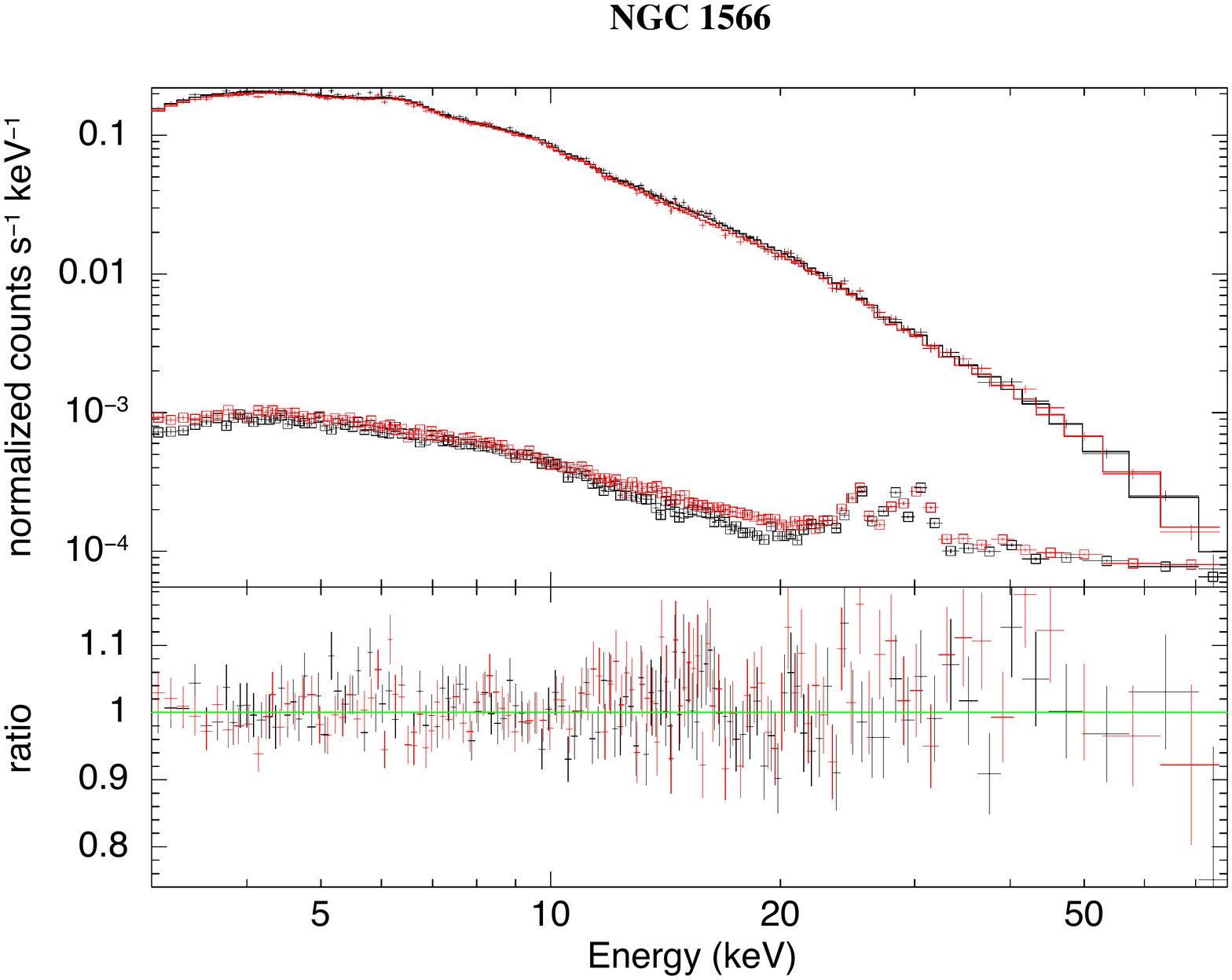}}
\subfloat{\includegraphics[width=0.33\textwidth]{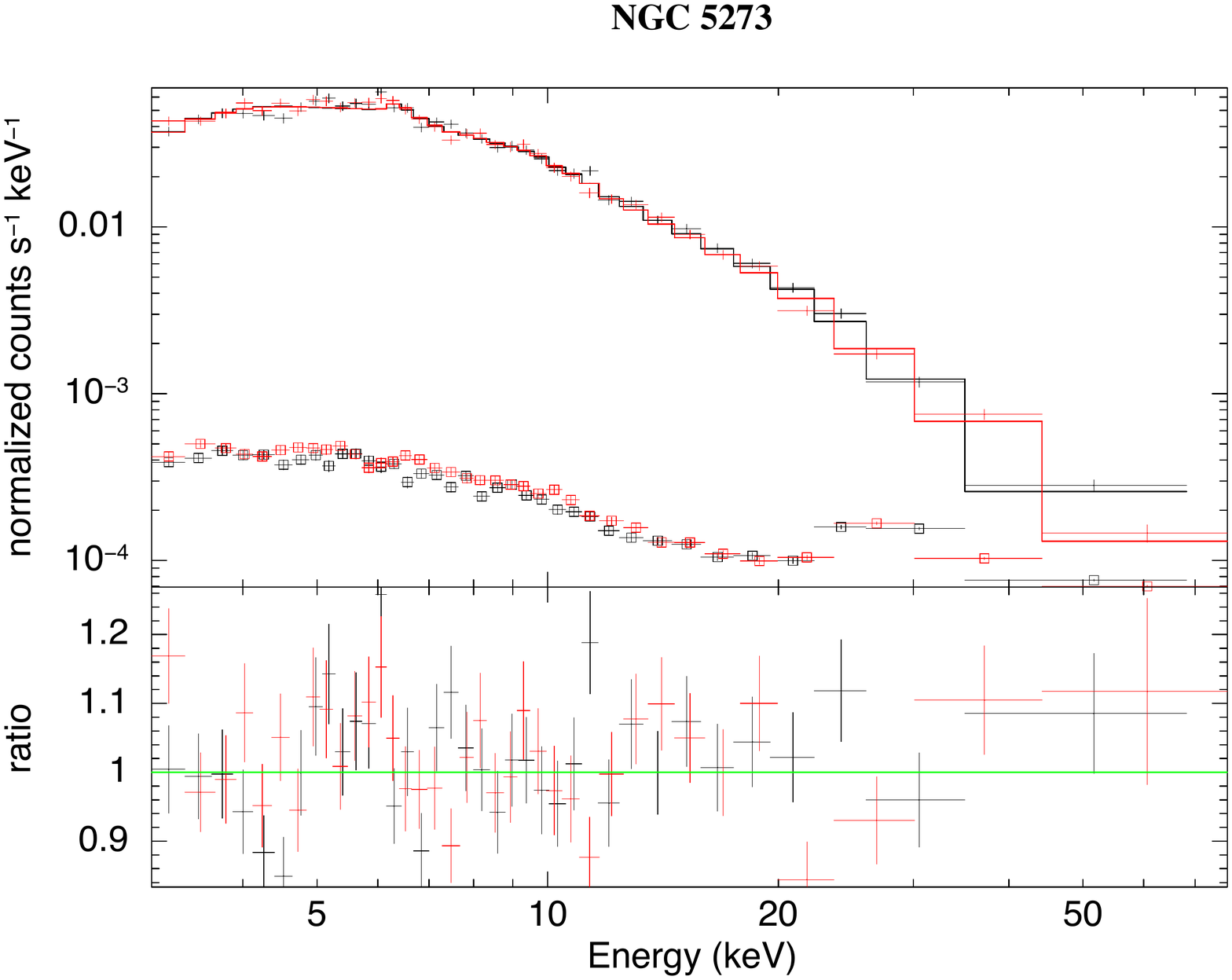}}\vspace{-6mm}\\
\subfloat{\includegraphics[width=0.33\textwidth]{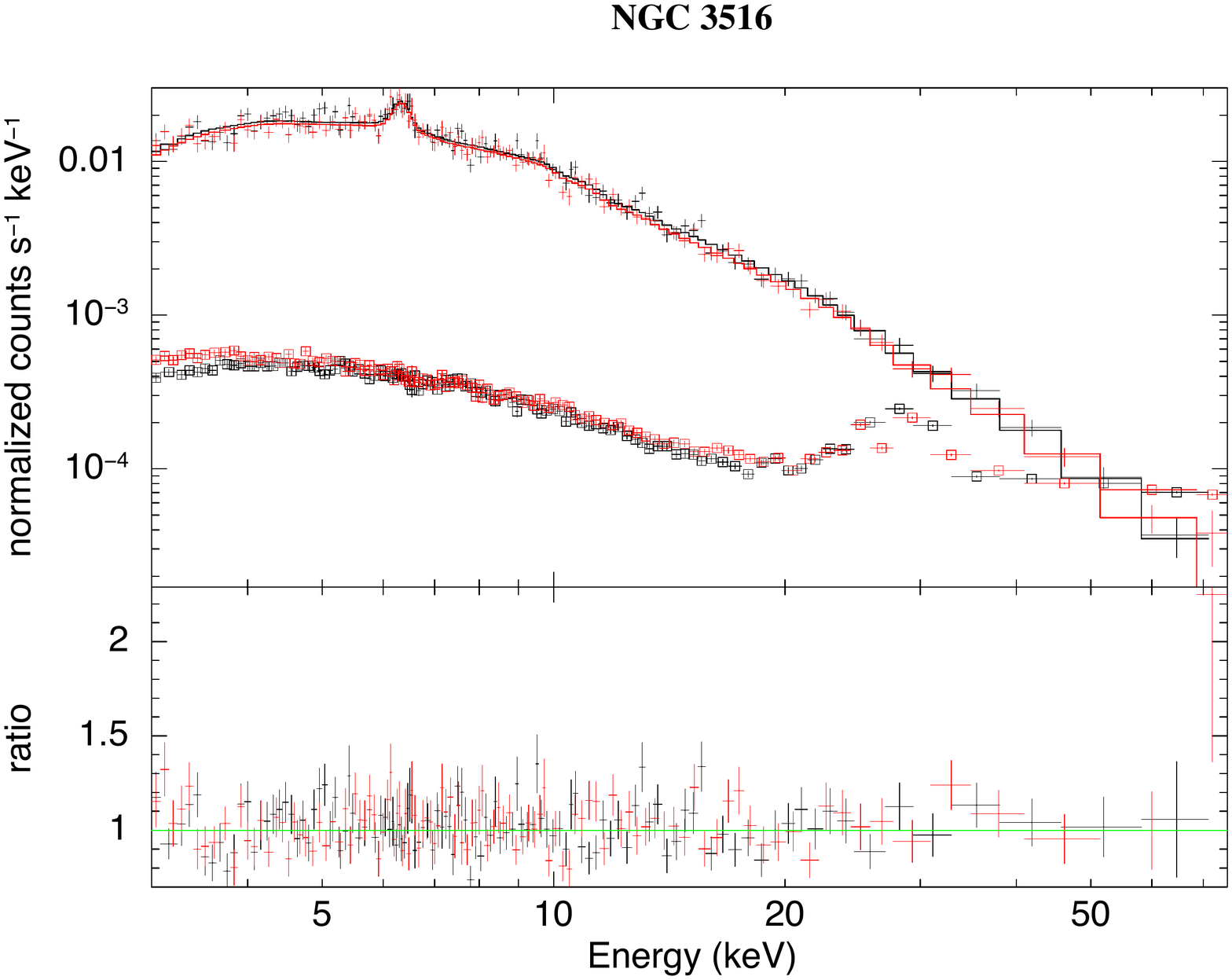}}
\subfloat{\includegraphics[width=0.33\textwidth]{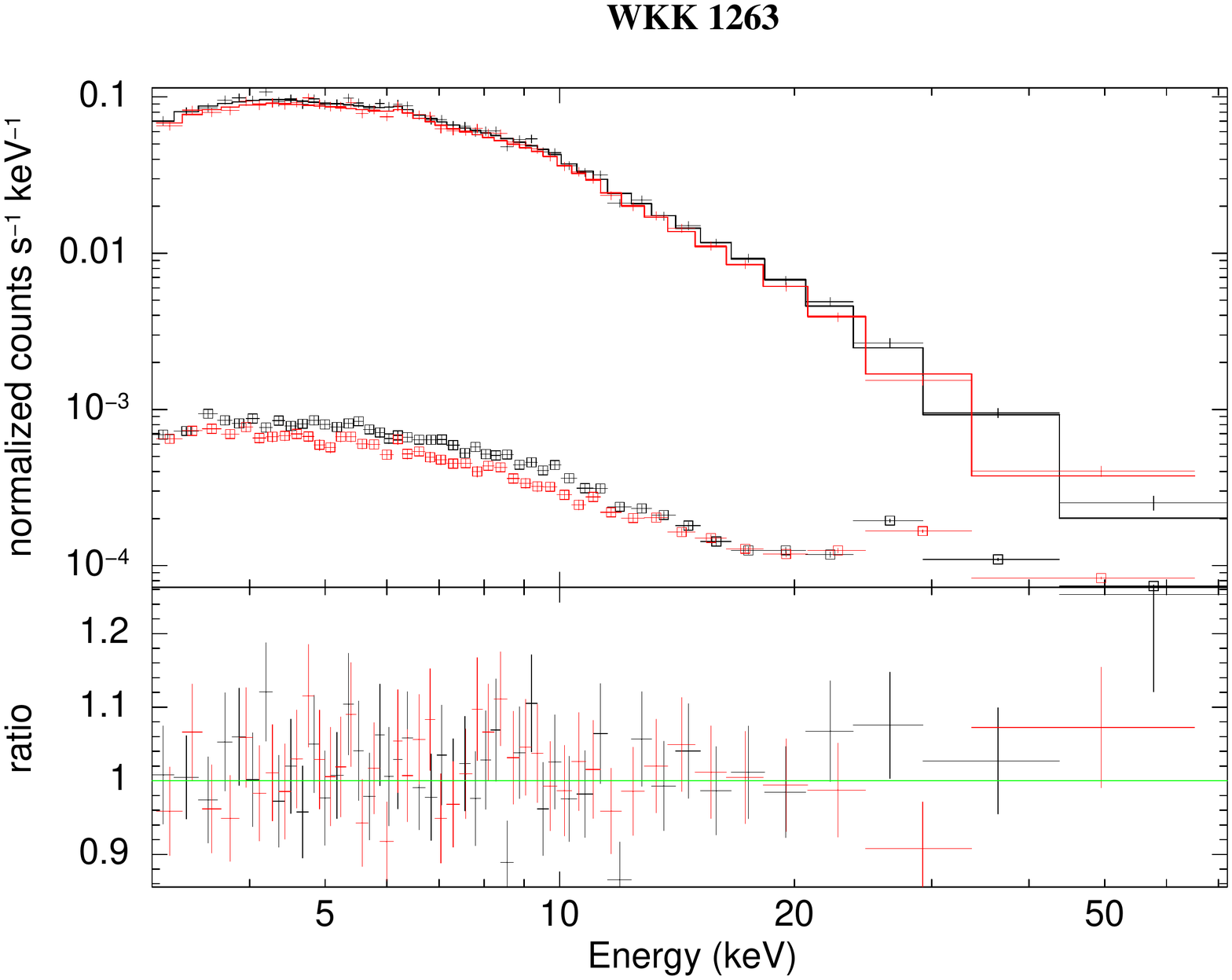}}
\subfloat{\includegraphics[width=0.33\textwidth]{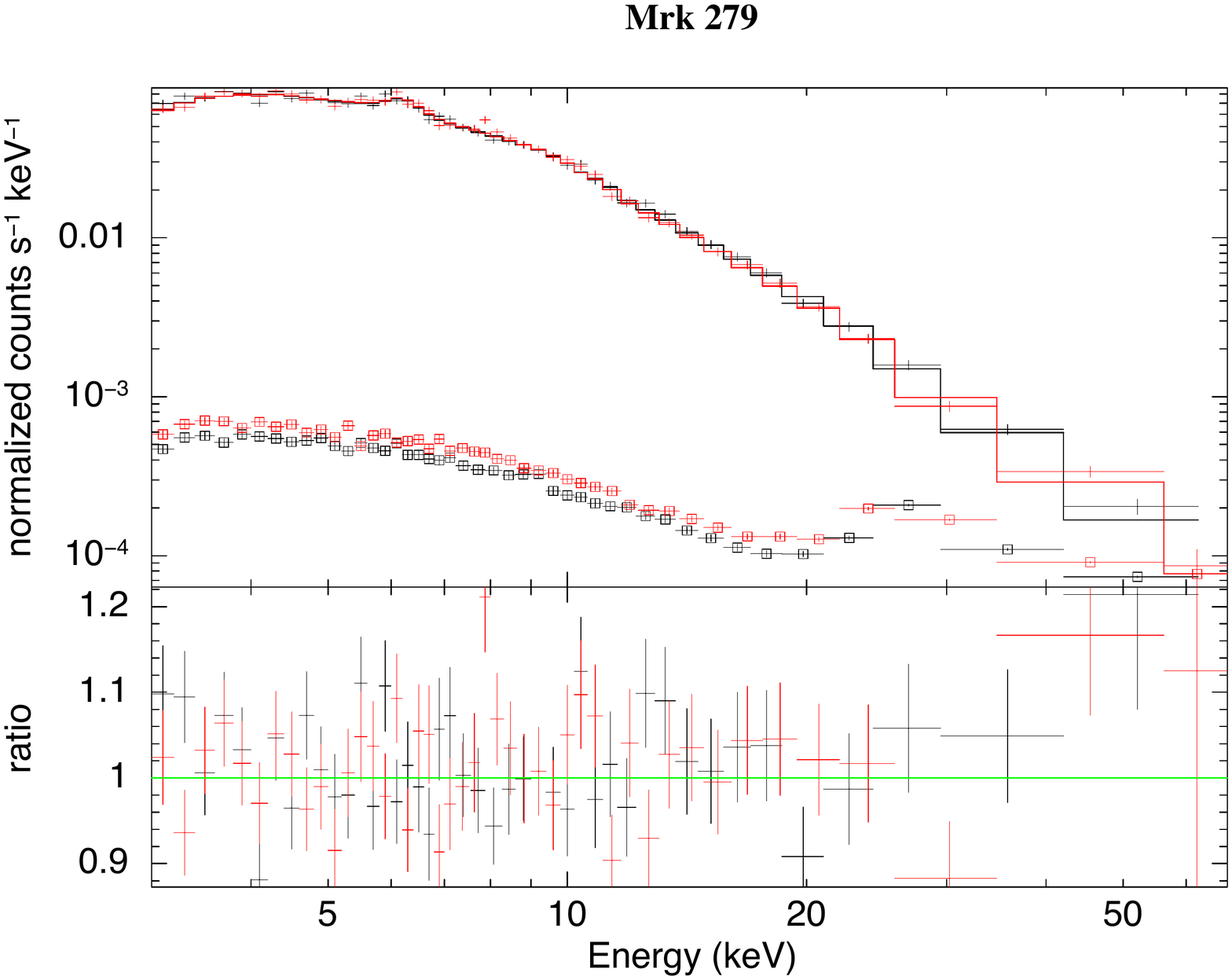}}\vspace{-6mm}\\
\subfloat{\includegraphics[width=0.33\textwidth]{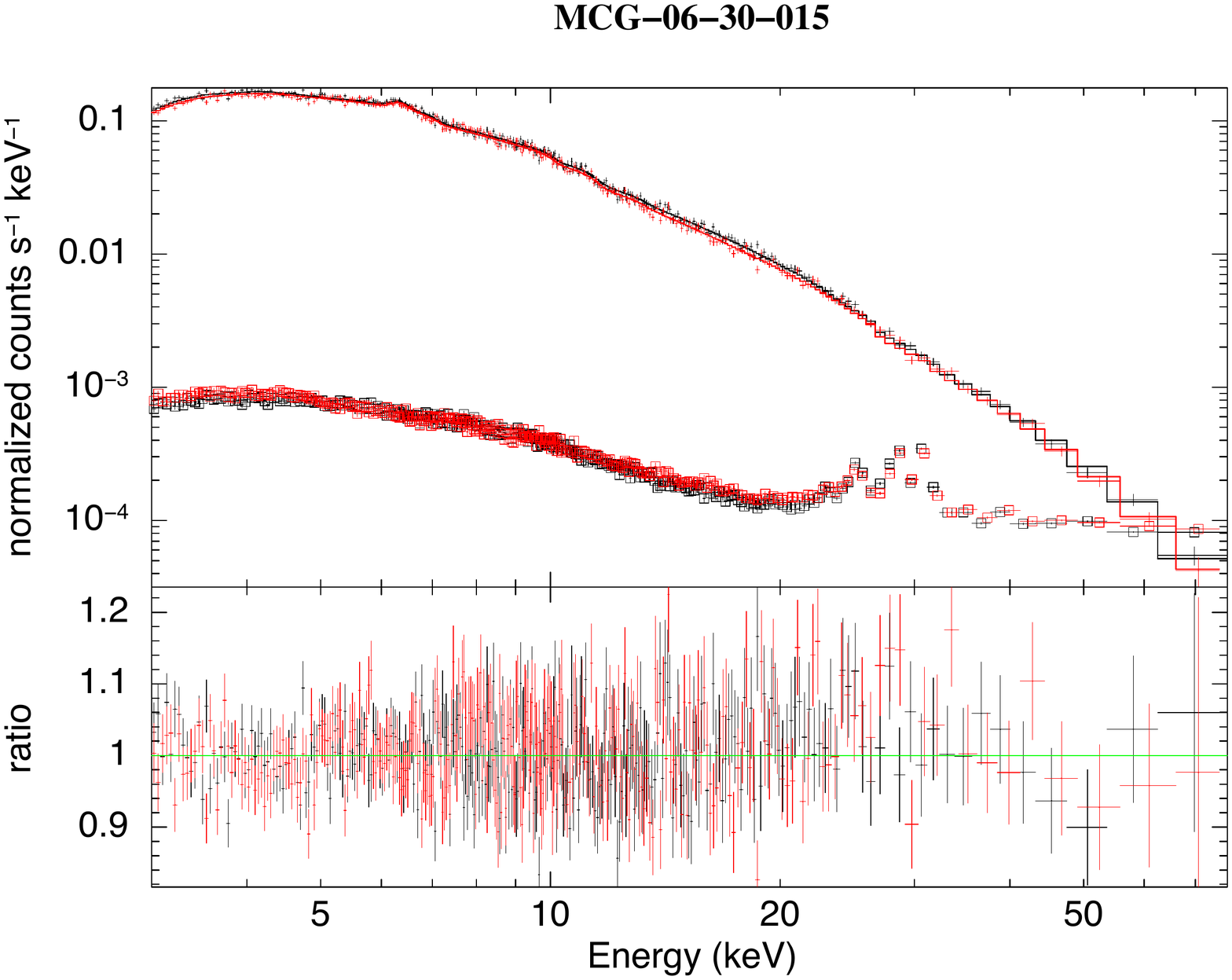}}
\subfloat{\includegraphics[width=0.33\textwidth]{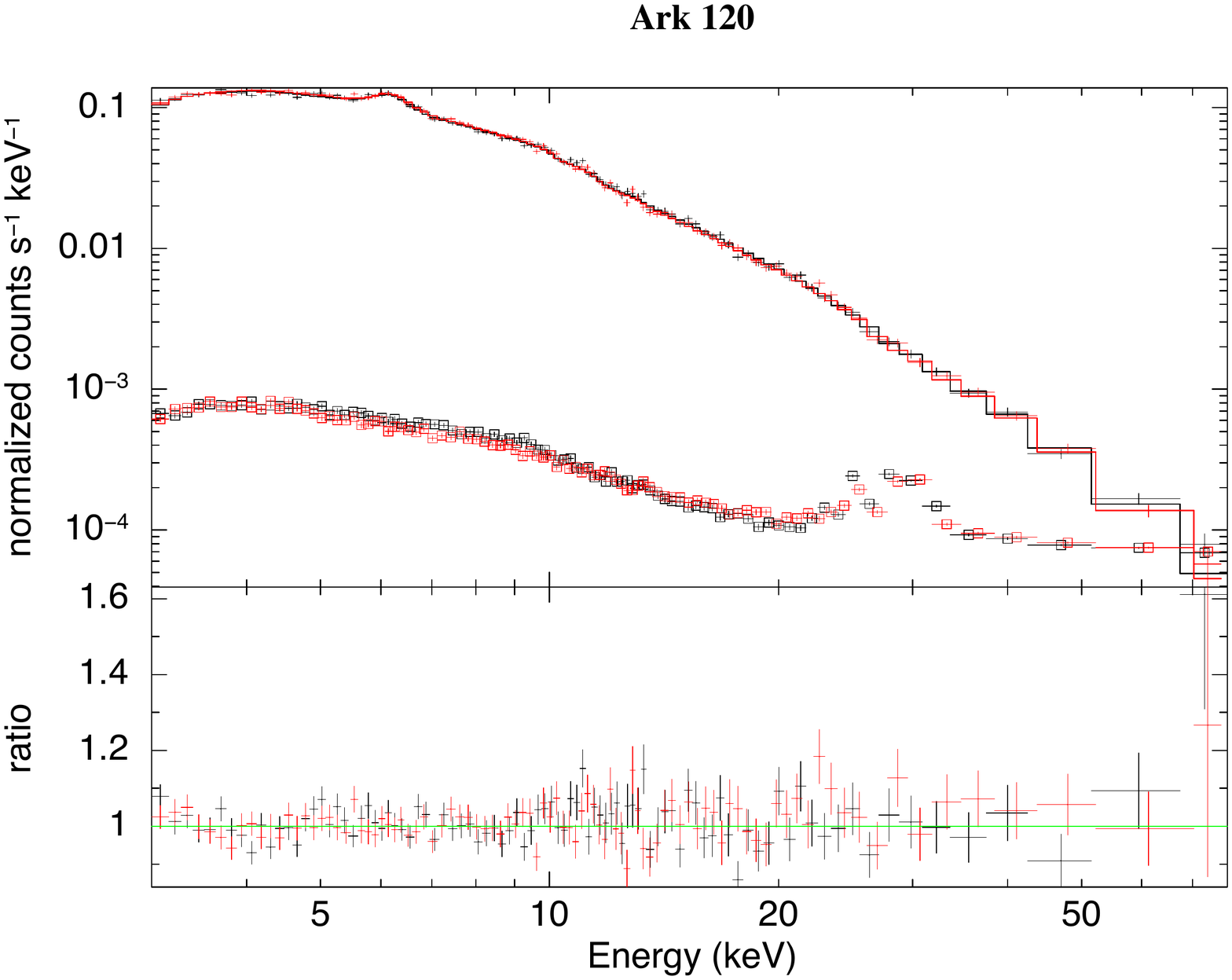}}
\subfloat{\includegraphics[width=0.33\textwidth]{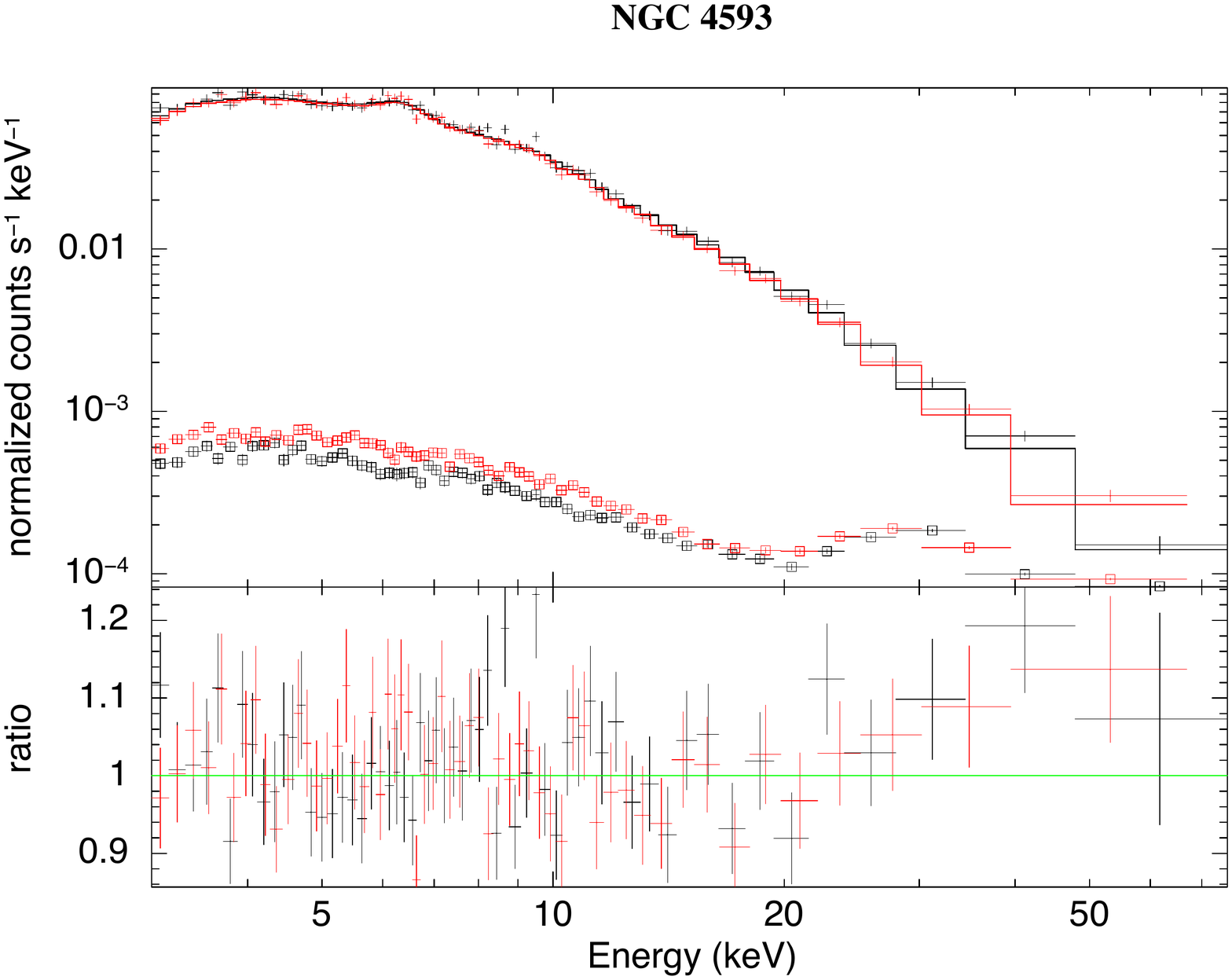}}\vspace{-6mm}\\
\subfloat{\includegraphics[width=0.33\textwidth]{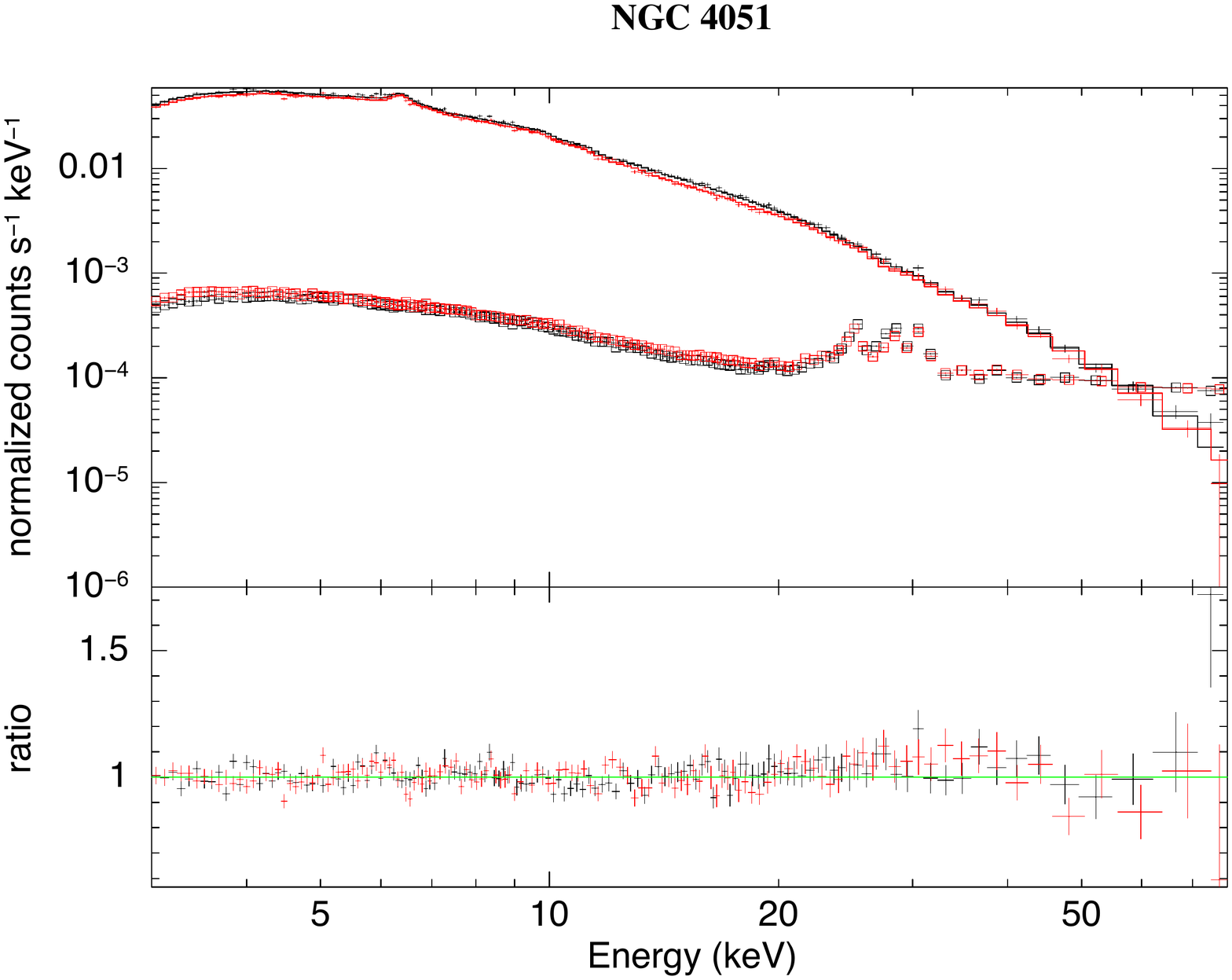}}
\caption{NuSTAR source and background spectra (estimated by NUSKYBGD), the best-fit models and the data-to-model residual ($pexrav$) ratios of the 10 sources with large \ec lower limits ($>$ 400 keV), ordered from low to high by the \ec lower limit). Note in only a few of them, e.g., NGC 3516 and NGC 4051, the spectra appear background dominated (i.e., the background fluxes larger than source fluxes) at above 50 keV.
Spectra from both FPMA (black) and FPMB (red) modules are given and further rebinned for visualization purposes.}
\label{fig:spectra}
\end{figure*}  

\clearpage

\bibliography{RLRQ}{}
\bibliographystyle{aasjournal}

\end{document}